\documentclass[namedreferences]{solarphysics}

\usepackage[optionalrh]{spr-sola-addons} 
\usepackage{graphicx}        		
\usepackage{amssymb}        		
\usepackage{color}           		
\usepackage{url}             		
\usepackage{booktabs}			
\usepackage[pdfborder={0 0 0 },urlcolor=blue,breaklinks]{hyperref}
\usepackage{lineno,xcolor}

\ifx \doiurl    \undefined \def \doiurl#1{\href{http://dx.doi.org/#1}{\textsf{doi}}}\fi
\ifx \adsurl    \undefined \def \adsurl#1{\href{http://adsabs.harvard.edu/abs/#1}{\textsf{ADS}}}\fi
\ifx \arxivurl  \undefined \def \arxivurl#1{\href{http://arxiv.org/abs/#1}{\textsf{arXiv}}}\fi
\newcommand{\bb}{\vec B}  
\newcommand{\rr}{ \vec r}
\newcommand{\xx}{ \vec x}
\newcommand{\bbb}{\vec b} 
\newcommand{\la}{\lambda} 
\newcommand{\si}{\sigma}  
\newcommand{\be}{\beta}   

\renewcommand{\vec}[1]{{\mathbfit #1}}

\begin{document}

\begin{article}

\begin{opening}

\title{Characterization of the Turbulent Magnetic Integral Length in the Solar Wind: 
From 0.3 to 5 Astronomical Units}
\author{M.E.~\surname{Ruiz}$^{1,2}$\sep
        S.~\surname{Dasso}$^{1,2,3}$\sep
        W.H.~\surname{Matthaeus}$^{4}$\sep
        J.M.~\surname{Weygand}$^{5}$   
       }
\runningauthor{M.E. Ruiz {\it et al.}}
\runningtitle{Turbulent Integral Length in the Solar Wind}
\institute{
$^{1}$ Instituto de Astronom\' ia y F\' isica del Espacio (CONICET-UBA), 
CC 67, Suc. 28, 1428, Buenos Aires, Argentina.\\
email: \url{meruiz@iafe.uba.ar} 
email:  \href{meruiz@iafe.uba.ar}{meruiz@iafe.uba.ar}\\
$^{2}$ Departamento de F\' isica, Facultad de Ciencias Exactas y Naturales, 
UBA, Pabell\'on 1 (1428), Buenos Aires, Argentina. \\
email:  \href{dasso@df.uba.ar}{dasso@df.uba.ar}\\
$^{3}$ Departamento de Ciencias de la Atm\'osfera y los Oc\'eanos, Facultad de Ciencias Exactas y Naturales, 
UBA, Pabell\'on 2 (1428), Buenos Aires, Argentina. \\
email:  \href{sdasso@at.fcen.uba.ar}{sdasso@at.fcen.uba.ar}\\
$^{4}$ Bartol Research Institute, Department of Physics and Astronomy, 
University of Delaware, Newark, DE, USA.\\
$^{5}$ Institute of Geophysics and Planetary Physics, University of 
California, Los Angeles, CA, USA.
           }
\begin{abstract}

The solar wind is a structured and complex system, in which the fields
vary strongly over a wide range of spatial and temporal scales. As an 
example, the turbulent activity in the wind affects the evolution in the 
heliosphere of the integral turbulent scale or correlation length [$\la$], 
usually associated with the breakpoint in the turbulent-energy spectrum 
that separates the inertial range from the injection range. This large 
variability of the fields demands a statistical description of the solar 
wind. In this work, we study the probability distribution function (PDF) 
of the magnetic autocorrelation lengths observed in the solar wind at 
different distances from the Sun. We use observations from {\it Helios}, 
ACE, and {\it Ulysses} spacecraft. We distinguish between the usual solar 
wind and one of its transient components (Interplanetary Coronal Mass Ejections, 
ICMEs), and study also solar wind samples with low and high proton beta [$\be_{\mathrm p}$]. 
We find that in the last 3 regimes the PDF of $\la$ is a log-normal function, 
consistent with the multiplicative and non-linear processes that take place 
in the solar wind, the initial $\la$ (before the Alfv\'enic point) being 
larger in ICMEs.
\end{abstract}

\end{opening}

\section{Introduction} \label{S_Intro} 

The solar wind (SW) is a very complex and structured system, where the 
fields are highly variable over different temporal and spatial scales. 
However, despite its complexity, different types of phenomena generally 
associated with different scales in the SW can be identified. 

At the global scale, the SW steady expansion has direct consequences on 
the typical length scales at which the bulk physical quantities that 
characterize the state of the system vary ({\it e.g.}, mass density, 
magnetic-field components, temperature). 
Between 0.3 astronomical unit (AU) and 5 AU from the Sun, these quantities 
typically decay as a power law with a negative exponent of the order of one to three
\cite{Mariani_libro_marron}.
Then, at a distance $D$ from the Sun, the \textquotedblleft steady 
expansion\textquotedblright~typical length scale can be estimated as $\approx D$.

Furthermore, different transient phenomena with origin at the solar surface 
produce disturbances to the steady SW. An example of these SW \textquotedblleft
transient structures\textquotedblright~is the phenomenon of fast transient streams 
of plasma from coronal holes \cite{Coronal_holes} or interplanetary coronal mass 
ejections (ICMEs), which have a magnetic topology radically different from the 
steady SW ({\it e.g.}, \opencite{Dasso_rev_MC_2005AdSpR}). These composite structures 
(which can contain several smaller sub-structures such as shock waves, plasma sheaths, 
{\it etc.}) are meso-scale objects in the system, with a range of sizes that are some 
fraction of $D$.

In SW turbulence, the largest spatial scale of the inertial range can be 
approximated by the turbulent integral scale [$\la$] (see Equation 
(\ref{E_def_lambda}) for a proper definition), which is also a proxy for 
the typical size of the \textquotedblleft energy-containing 
eddies\textquotedblright~({\it e.g.}, \opencite{Bill_JGR94_evol_turb_eddies}).
The inertial range extends from $\la$ to much smaller scales, involving 
turbulent processes along several orders of magnitude. It is very rich in 
non-linear processes (see for example \opencite{Coleman68}), combined with 
an important level of wave activity (see for example \opencite{BelcherDavis71}).
This complex turbulent activity affects the evolution of different aspects
of the SW fluctuations, such as the fluctuating intensity, the integral length 
[$\lambda$], the level of Alfv\'enicity \cite{Tu-Marsch-1995}, anisotropy 
\cite{Matthaeus90,dasso05,Ruiz2011}, {\it etc}.
In particular, it is well known that $\la$ increases with heliocentric distance 
\cite{Tu-Marsch-1995}. Near Earth $\lambda_{{\mathrm 1 AU}}$ is $\approx 0.0079$ AU 
\cite{Matthaeus05_msc} while $\lambda_{{\mathrm 10 AU}}$ is $\approx 0.046$ AU in the 
SW near Saturn \cite{Smith2001JGR}.

All of these physical phenomena, associated with significantly different spatial scales, 
are coupled. For instance: 
i) the decay of the total solar-wind pressure (determined by its \textquotedblleft steady 
expansion\textquotedblright~scale) plays the major role during the long-term interaction 
between magnetic clouds and their environment \cite{Pascal-Dasso,Adri_inner,Adri_outer}, 
ii) the presence of shear in the velocity profile ({\it e.g.}, associated with CIRs or ICMEs) 
can produce instabilities and introduce energy into the outer scales of the turbulent 
inertial range \cite{Goldstein_MHD_rev_95},
iii) turbulent properties control the drag on ICMEs and many other large-scale processes 
\cite{bill_velli_who_needs_t}, {\it etc}.

An important entity for studying fluctuations of turbulent fields is the autocorrelation 
function. For the magnetic field, the average trace of the two-point/two-time correlation 
tensor is
\begin{equation}\label{E_R_def}
 R([\xx,t],[\rr, \tau])=\langle \bbb(\xx,t)\cdot \bbb(\xx + \rr,t + \tau) \rangle
\end{equation}
where $\bbb$ is the fluctuating component of $\bb$ and $[\rr,\tau]$ are the spatial and temporal 
lags, respectively. We can drop the $[\xx,t]$ dependence in Equation (\ref{E_R_def}) if we assume 
stationarity and homogeneity of the medium \cite{1982Matthaeus,2013LivRev}.
Further, we may assume the Taylor frozen-in-flow hypothesis \cite{Taylor-1938} to be valid in the 
supersonic and super-Alfv\'enic SW; that is, the fluctuating fields are convected past the spacecraft 
in a shorter time than their characteristic dynamical timescale. Then we can ignore the intrinsic 
temporal dependence of the fluctuations in Equation (\ref{E_R_def}), resulting in
\begin{equation}\label{E_R_taylorhyp}
 R(\rr)=\langle \bbb(\vec 0)\cdot \bbb(\rr) \rangle
\end{equation}
The spatial decorrelation of the turbulence can be characterized
by the correlation length or integral scale
\begin{equation}\label{E_def_lambda}
\la=\frac{\int_0^{\infty}\langle\,\bbb(\vec 0)\cdot\bbb(\rr)\,\rangle}{\langle\bbb^2 \rangle {\mathrm d}r}
\end{equation}
Conventionally, this typical length-scale is understood as being a measure of 
the size of the turbulent energy-containing eddies in the flow \cite{Batchelor_turb}. 
Moreover, $\la$ can be linked to the scale associated with the spectral break 
that separates the injection range (meso-scales) from the inertial range: 
$\la$ can be seen as a kind of spatial frontier between the two domains.

Any description of the complex SW physical system should be complemented 
by a statistical description of the fields, since important information 
about turbulent systems resides at a statistical level and, to this day, 
it is not possible to measure initial or boundary conditions \cite{burlaga2000}. 

Log-normal distributions are frequent in nature across the different branches 
of science \cite{Limpert01}, and are believed to be a consequence of multiplicative 
processes ({\it e.g.}, \opencite{Montroll}). In particular, in the field of space and 
solar physics, many authors have considered log-normal distributions when modelling 
quantities of interest such as the Dst index \cite{Campbell_Dst_Logn}, the magnitude 
of the magnetic field fluctuations \cite{Burlaga_lognB,PadhyeJGR01}, SW speed, proton 
density and temperature \cite{burlaga2000}, proton plasma beta and Alfv\'en speed 
\cite{Mullan2006}. 

As far as we know, the probability distribution functions (PDFs) of autocorrelation 
lengths [$\la$] of the solar-wind fluctuating magnetic field have not been studied. 
\inlinecite{Wicks2010SoPh} reported an asymmetric shape for the observed PDF of the 
correlation lengths of the magnetic field magnitude at 1 AU. \inlinecite{Matthaeus_PRL86} 
had theoretically postulated that $\la$ is log-normally distributed. The authors 
explained that the structures that initiate the cascade in the inertial range, amplify 
their initial size $\la_0$ during their transport into the SW from the solar surface, 
employing a mechanism of successive magnetic reconnection events to increase 
the size of magnetic structures. 
This occurs $M$ times each one by a factor $(1+\epsilon)$ yielding a final size given
by $\la=\la_0(1+\epsilon)^M$ with $\la$ the correlation length of the fluctuations. 
If $M$ is sufficiently large, the random variable $\ln(\la)$ will be normally
distributed and therefore $\la$ log-normally distributed.

Thus, the discussion presented in this Section motivates us to study 
$\la$ in the SW and its evolution. One of the main aims of 
this article is to provide an observational characterization of the PDF of $\la$.

\section{Data and Procedure}\label{S_Data}

We use the magnetic field and plasma observations collected by different spacecraft 
that repeatedly explored the inner and outer heliosphere at different heliocentric 
distances $D$. In particular we analysed \textit{in-situ} solar wind observations 
from the following four probes: {\it Helios} 1 (H1), {\it Helios} 2 (H2), Advanced 
Composition Explorer (ACE), and {\it Ulysses}. Our H1 and H2 time series cover the 
period from December 1974 to June 1981, the temporal cadence is 40~seconds and 
observations are essentially on the Ecliptic plane. Near-Earth observations coming 
from the ACE spacecraft cover the period from February 1998 to March 2008, with time 
cadence of one minute. {\it Ulysses} series range from November 1990 to May 2009, and the 
temporal cadence is of one minute. We restrict our {\it Ulysses} observations to the 
Ecliptic plane by choosing heliocentric latitudes [$\theta$] such that $|\theta|<$30$^{\circ}$. 

For each spacecraft (labeled $s$) we group the data into 24-hour-length intervals [$I$], 
thus obtaining $N^{s}_{1}$ subseries (intervals). Due to the presence of many gaps in the 
{\it Helios} data, we repeat the procedure for H1 and H2 only by shifting the data by 12 
hours to obtain $N^{s=H1,H2}_{2}$ additional intervals, thus maximizing {\it Helios} data 
utilization.

We avoid samples with very low statistical significance by retaining only those intervals 
encompassing at least the 30$\,\%$ of the observations expected for the cadence mentioned 
for each spacecraft.

We compute the magnetic correlation functions and respective correlation lengths as follows:
in each interval [$I$] and for each spacecraft [$s$], from the observed magnetic field time 
series [${\bf B}^{I,s}$] we construct the magnetic fluctuations as 
${\bf b}^{I,s}={\bf B}^{I,s}-{\bf B}^{I,s}_{0}$, with ${\bf B}^{I,s}_{0}$ a linear fit to 
${\bf B}^{I,s}$ data. Here we identify the fitted field [${\bf B}^{I,s}_{0}$] with the local 
(within the interval) estimate of the average magnetic field in an ensemble.

Next, we compute each correlation function [$R^{I,s}$] using the Blackman--Tukey technique, in 
the same way as was done in \inlinecite{Milano2004}. Inside the Alfv\'enic point, the different 
initial conditions will yield different initial values for $R({\bf 0})=\langle b^2\rangle$ and 
for $\la$. So, in order to be able to compare intervals with different fluctuating amplitude, we 
normalize the correlation functions as $R^{I,s}_{norm}=R^{I,s}/R( {\bf 0})^{I,s}$. For simplicity 
of notation, we drop the labels \textit{norm} and $s$ hereafter.

Figure 1 in \inlinecite{Ruiz2011} shows a typical correlation function in the inner heliosphere 
that can be obtained with the Blackman--Tukey technique. Correlation functions in the outer 
heliosphere have a similar shape.

A simple approximation that is often used to the shape of $R^{I}$, at large scales and in the 
long wavelength part of the inertial range, is an exponential decay $R\approx\exp(-r/\la)$. This 
approximation provides us with two methods, i) and ii), to estimate the magnetic autocorrelation 
length $\la^{I}$ in each interval.
The first method determines an estimate of $\lambda^{I}_{i}$ as the value of the spatial lag such 
that the decreasing function $R^{I}$ reaches $\exp(-1)$ for the first time, {\it i.e.}, $R^{I}(\lambda^{I}_{i})=1/e$. 
Method ii) consists in parametrizing the correlation function as 
$\ln(R)\approx -r/\lambda$. We estimate $\lambda^{I}_{ii}$ as minus the inverse of the slope 
obtained from a linear fit to $\ln(R)$ vs. $r$.

It is accepted that for steady turbulence, magnetic autocorrelation functions behave as shown in
Figure 1 in \inlinecite{Ruiz2011}. Departures from this shape can imply the presence of transient 
events ({\it e.g.}, large-scale current-sheet crossings).
 
Under the approximation $R\approx\exp(-r/\la)$ for the autocorrelation functions, we find that 
steady turbulent intervals are characterized by $\lambda_{i}^{I} \simeq \lambda_{ii}^{I}$, 
while intervals far away from steady turbulence show values of $\lambda_{ii}^{I}$ very different 
to $\lambda_{i}^{I}$. This fact allows us to define a quality factor $F$ of the correlation 
function, based on the two estimates of the correlation length. We define this factor as
\begin{equation}
 F^{s,I}=\frac{\lambda_{ii}^{s,I}-\lambda_{i}^{s,I}}{\lambda_{ii}^{s,I}+\lambda_{i}^{s,I}}
\nonumber
\end{equation}
Note that a small positive offset in $F$ is expected due to the systematic differences in 
method i) and ii) (see \inlinecite{Ruiz2011}).
\begin{figure}
\centerline{
\includegraphics[width=0.334\textwidth,height=.334\textwidth]
{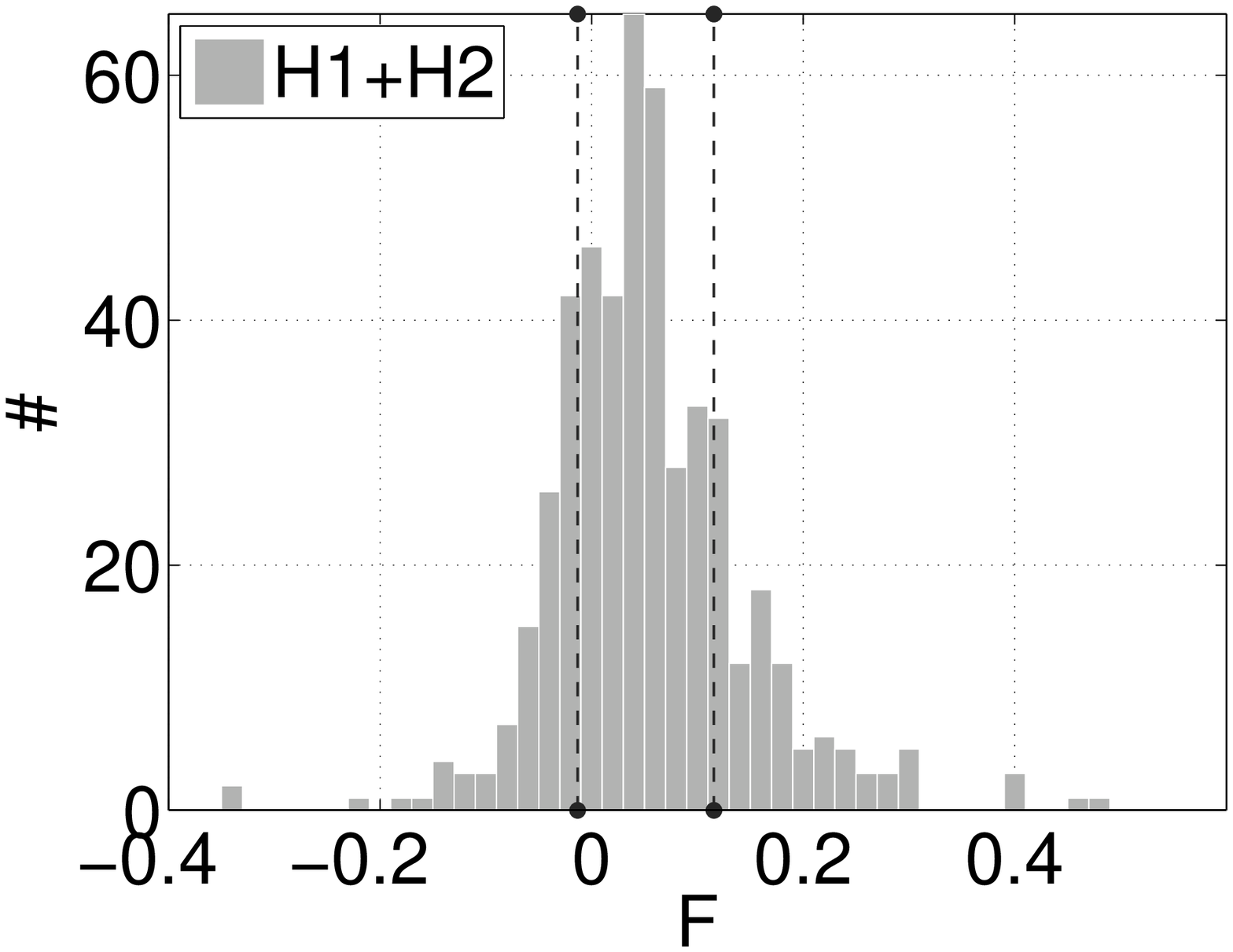}
\includegraphics[width=0.334\textwidth,height=.334\textwidth]
{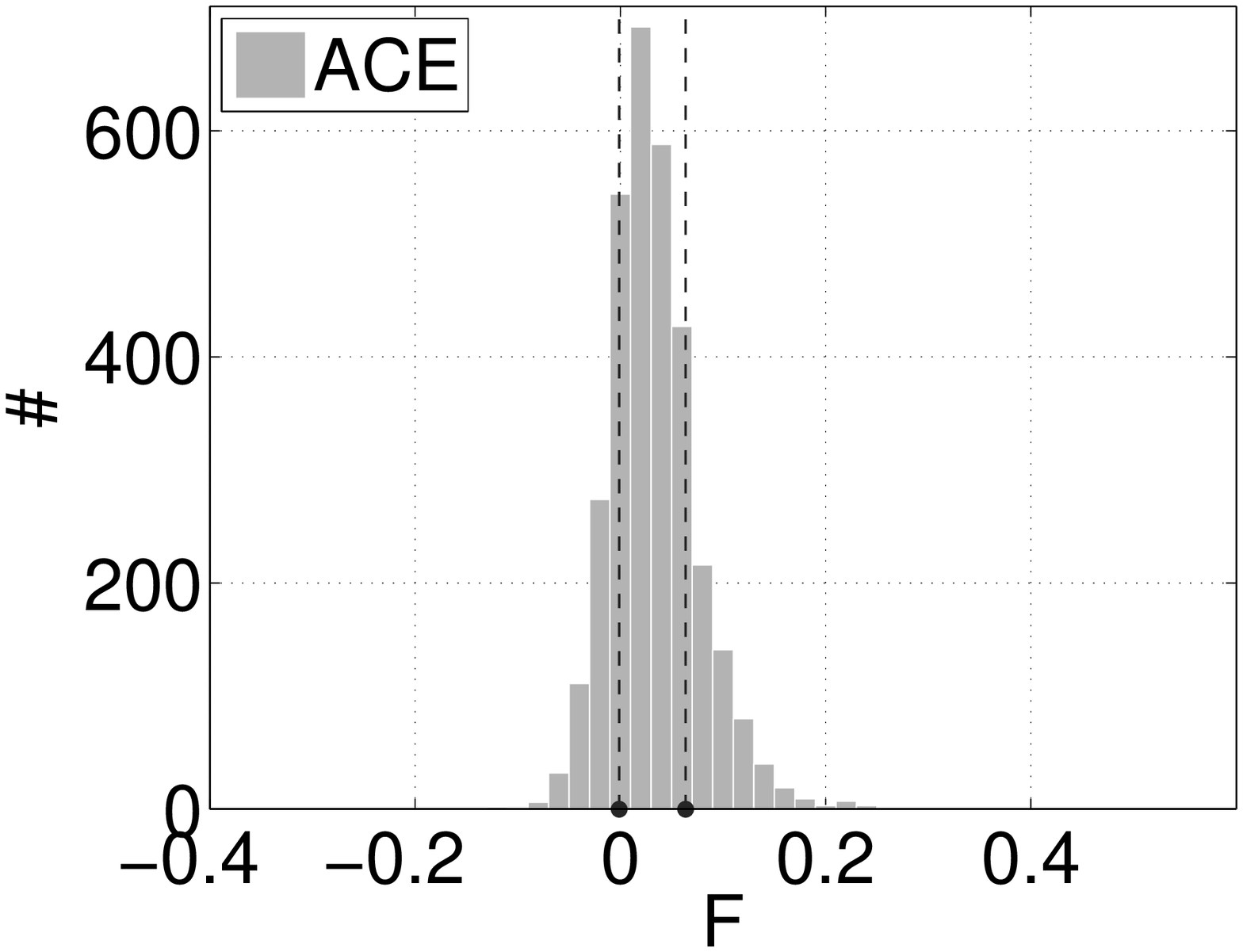}
\includegraphics[width=0.334\textwidth,height=.334\textwidth]
{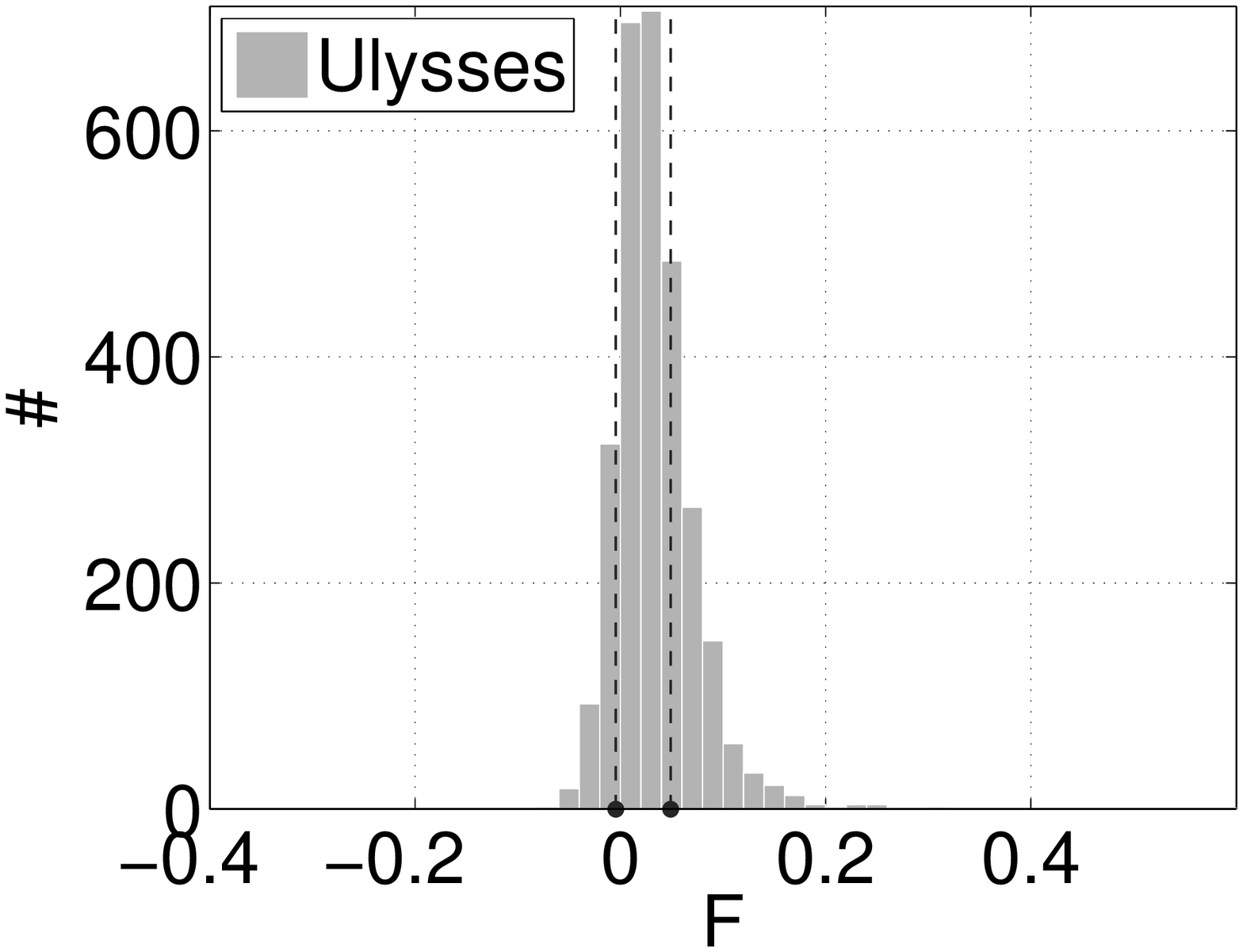}}
\vspace{-0.00\textwidth}
\centerline{\Large \bf
\hspace{0.138\textwidth}  \color{black}{\small{(a)}}
\hspace{0.275\textwidth}  \color{black}{\small{(b)}}
\hspace{0.268\textwidth}  \color{black}{\small{(c)}}
\hfill}
\vspace{0.030\textwidth}    
\caption{Histograms of the quality factor index [$F$]. Dashed vertical lines indicate
the position of the 20$^{\mathrm{th}}$ (left) and 80$^{\mathrm{th}}$ (right) percentile. 
Panels (a), (b), and (c) correspond to H1$+$H2, ACE and {\it Ulysses} data respectively.}
\label{histo_F}
\end{figure}
%
Figure \ref{histo_F} shows the distribution of the quality factor index [$F$] for the three 
missions analyzed in this article. We select the best intervals based on the value of $F$ in 
each case, by retaining only those that fulfilled the following two conditions: $F$ values 
within the interval larger than the value of the 20$^{\mathrm{th}}$ percentile and those with 
$F$ values smaller than the value of the 80$^{\mathrm{th}}$ percentile. These intervals are 
the ones between the two dashed vertical lines in Figure \ref{histo_F}. 
Different ranges for $F$ were explored, arriving at qualitatively similar 
results. The main effect of modifying the $F$ ranges is to vary the intervals considered 
for each spacecraft.

Finally, after the different selection criteria, our collection of usable data includes 
$N^{s=U}_{1}=$1976 for {\it Ulysses},
$N^{s=A}_{1}=$1919 for ACE,
and 
$N^{s=H1+H2}_{1}+N^{s=H1+H2}_{2}=$846 for {\it Helios}, where $s=H1+H2$
indicates that we have gathered H1 and H2 intervals into only one data set representative of 
the inner heliosphere.

\section{Evolution of $\lambda$ with Heliocentric Distance and Nominal Aging}\label{S_evolution}
	
Turbulent structures evolve and are advected by the solar wind throughout the 
heliosphere with the SW velocity [$V_{\mathrm{sw}}$]. A key quantity for closing MHD 
turbulence models is the similarity scale, usually identified with the correlation
scale for the fluctuations. Observational insight into the evolution of correlation lengths 
is useful when comparing with numerical solutions or choosing proper boundary conditions.
\begin{figure}[!t]
\centering
\includegraphics[width=.49\textwidth,height=.45\textwidth]
{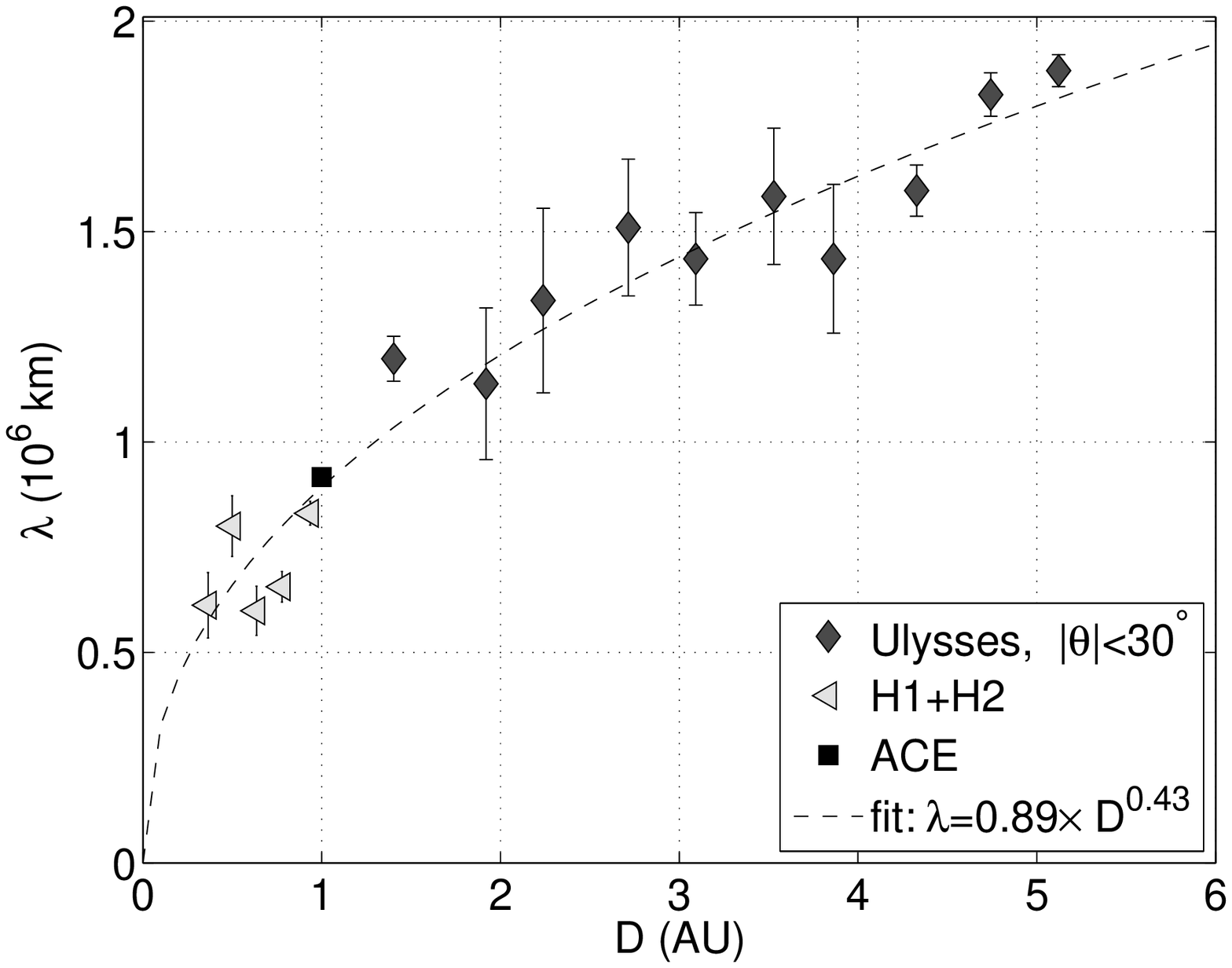}
\includegraphics[width=.49\textwidth,height=.45\textwidth]
{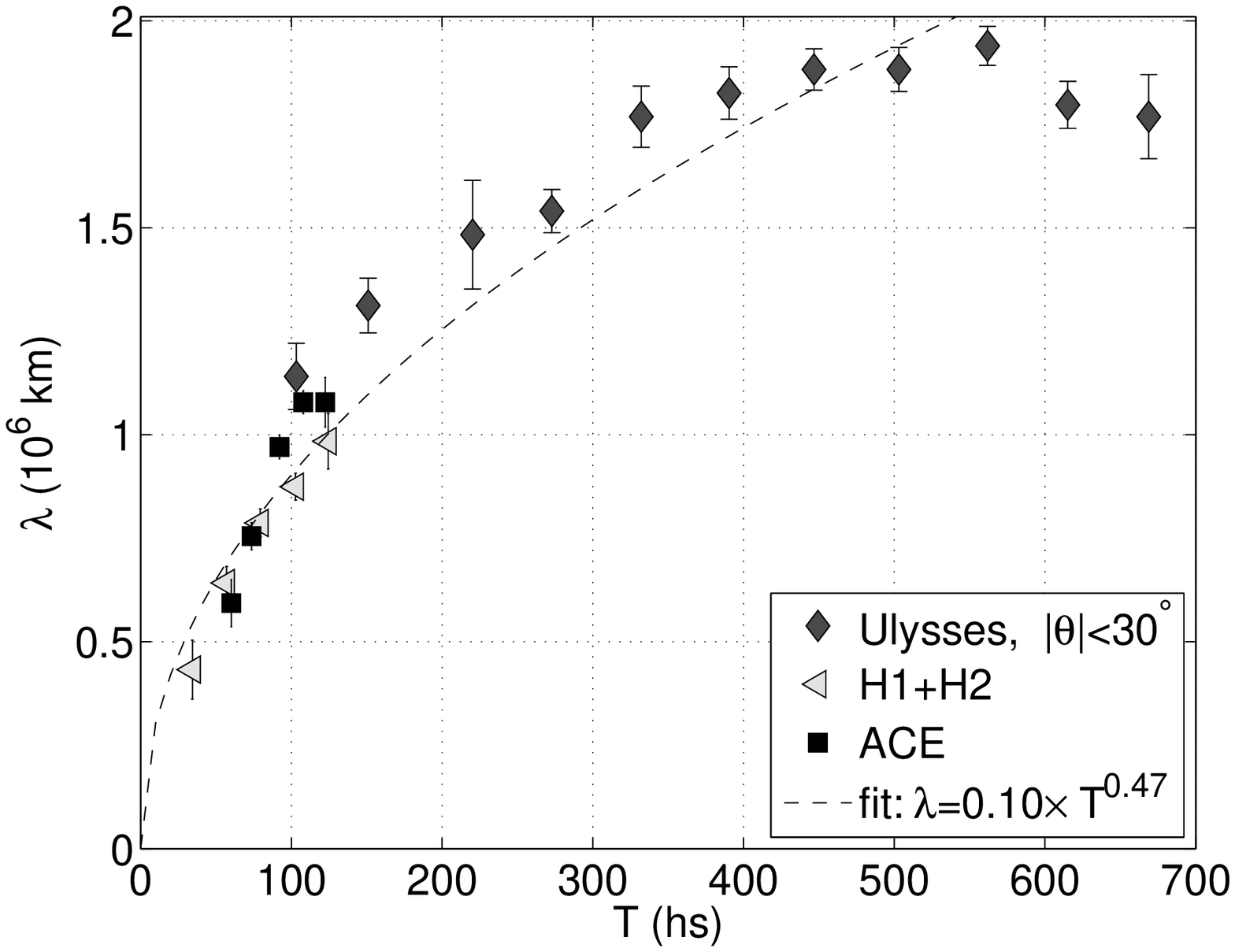}
\caption{Observed $\la$ vs. heliocentric distance [$D$] (left panel) and 
versus the solar-wind age (right panel). Bars show the error of the mean.}
\label{F_lambda_vs_dist_age}
\end{figure}

The left panel of Figure~\ref{F_lambda_vs_dist_age} shows how the observed $\la$ by {\it Helios} 
(triangles), ACE (squares), and {\it Ulysses} (diamonds) evolves with heliocentric distance. The 
observations have been grouped into bins of different width ($\Delta D=$0.14 AU for H1$+$H2 data 
and $\Delta D=$0.4 AU for {\it Ulysses} data), and each value of the vertical axis is the median 
of $\la$ within the bin. 

$\la$ increases with heliocentric distance both in the inner and outer heliosphere, as has been 
shown in previous observational works ({\it e.g.}, \opencite{Matthaeus_SW9}; \opencite{BrunoRev2005};
\opencite{Bruno_2009_EMP}; \opencite{DAmicis_2010ApJ}) and model calculations ({\it e.g.}, \opencite{Smith2001JGR}). 
A least-squares fit to the data (dashed line), illustrates this behavior, yielding a power law 
$\la(D)=0.89 (D/1\,AU)^{0.43}\times 10^6$\,km. Other exponents for the power laws have been reported 
({\it e.g.}, \opencite{KleinMatthaeus_SW7}). Moreover, between 1 AU and 5 AU, the growth rate of $\la$ 
with heliocentric distance is $\Delta \la/\Delta D \approx$ 0.0015, very close to the predictions of 
the model of \inlinecite{Smith2001JGR} with the strongest shear as the driver of the turbulence. On 
the contrary, observations reported by \inlinecite{DAmicis_2010ApJ}, show a larger growth rate between 
1 AU and 1.4 AU, $\Delta \la/\Delta D \approx$ 0.063 than observations reported here 
($\Delta \la/\Delta D \approx$ 0.0023). 	
Nevertheless, while they observed fast (Alfv\'enic) solar wind, we observed mixed fast and slow wind. 
Shear intensity is typically higher in slow than fast solar wind \cite{McComas2003GRL}, 
and the more intense the shear is, the more slowly correlation lengths increase. On the Ecliptic plane, 
the slow wind is more frequently encountered than the fast wind, so that slow SW properties 
are favored when computing averages. Other authors have studied the evolution of 
turbulent fluctuations in fast polar wind ({\it e.g.}, \opencite{Bavassano1982}; \opencite{Horbury_1995GRL};
\opencite{Horbury_1996AyA}). These authors showed that the breakpoint wavenumber in high latitude flows 
is smaller than that one in low latitude flows at similar heliocentric distances, revealing that polar 
fluctuations are less evolved than ecliptic fluctuations.
Finally, this growth of $\la$ is consistent with the shift towards low frequencies of the spectral 
break (this concept that was first introduced by \opencite{TUetal1984}) which separates the inertial 
range from the large-scale injection range, revealing that non-linear interactions at large heliocentric
distances are still taking place.

While traveling throughout the heliosphere, turbulent structures will reach 
a spacecraft located at $D$ after a time $\approx D/V_{\mathrm{sw}}$.
For each analyzed interval [$I$], we compute what we call the~\textquotedblleft 
age\textquotedblright~of the interval $I$: $T^{I} = D^{I}/V^{I}_{\mathrm{sw}}$. Then 
$T^{I}$ corresponds to the nominal time it takes a solar wind parcel 
[$I$] moving at speed $V_{\mathrm{sw}}^{I}$ to travel a given distance from the Sun 
to the spacecraft located at $D^{I}$.

The right panel of Figure~\ref{F_lambda_vs_dist_age} shows the evolution of $\la$ 
with $T$. The observations have been grouped into $T$-bins of different width
($\Delta T=$25 hours for H1$+$H2 data, $\Delta T=$18 hours for ACE and $\Delta T=$58 
hours for {\it Ulysses} data), and each value of the vertical axis is the median of $\la$ 
within the bin. 

Correlation lengths steadily grow with age until around $\approx$~500 
hours, but then they seem to decrease. A least-squares fit to the data (dashed 
line) yields a power law $\la(T)=0.10 (T/1\,\mathrm{hour})^{0.47}\times 10^6$\,km. This globally
increasing trend is consistent with numerical simulations derived from models for MHD 
turbulence based on the K\'arm\'an and Howarth HD approach \cite{Oughton2006PhPl}.

\section{Characterization of the PDF of $\la$ }
\label{S_charac_histos}

In this Section we characterize the distribution of correlation 
lengths of the solar wind magnetic fluctuations on the Ecliptic plane at 
three different distances from the Sun ({\it i.e.} three different stations).
To allow a clear distinction between the three data sets ({\it i.e.} between 
stations), we limit {\it Helios} observations to heliocentric distances between 0.3~AU 
and 0.7~AU, and {\it Ulysses} observations to heliocentric distances between 3~AU and 
5.3~AU. 

\begin{figure}[!t]
\centerline{
\includegraphics[width=0.333\textwidth,height=.333\textwidth]
{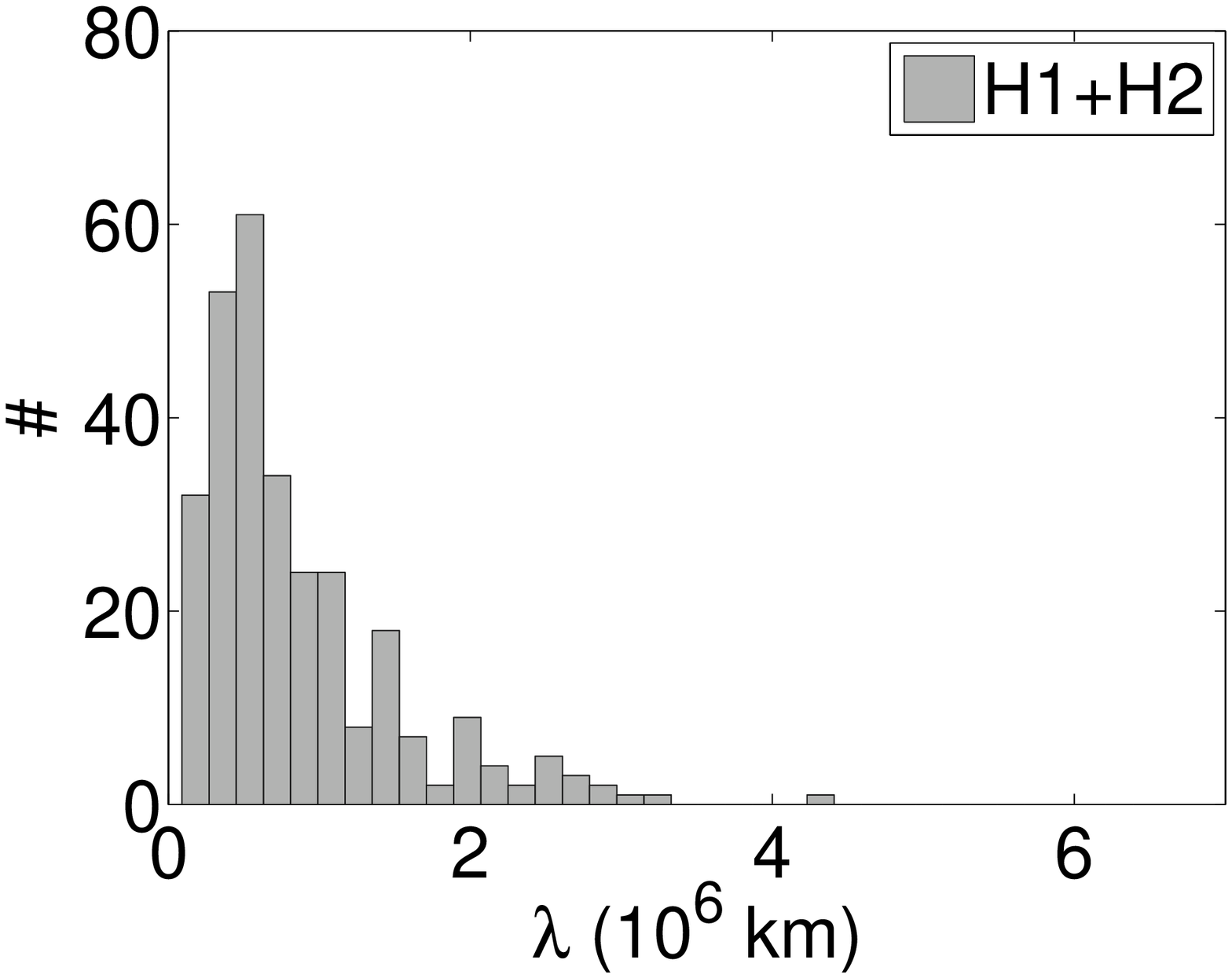}
\includegraphics[width=0.333\textwidth,height=.333\textwidth]
{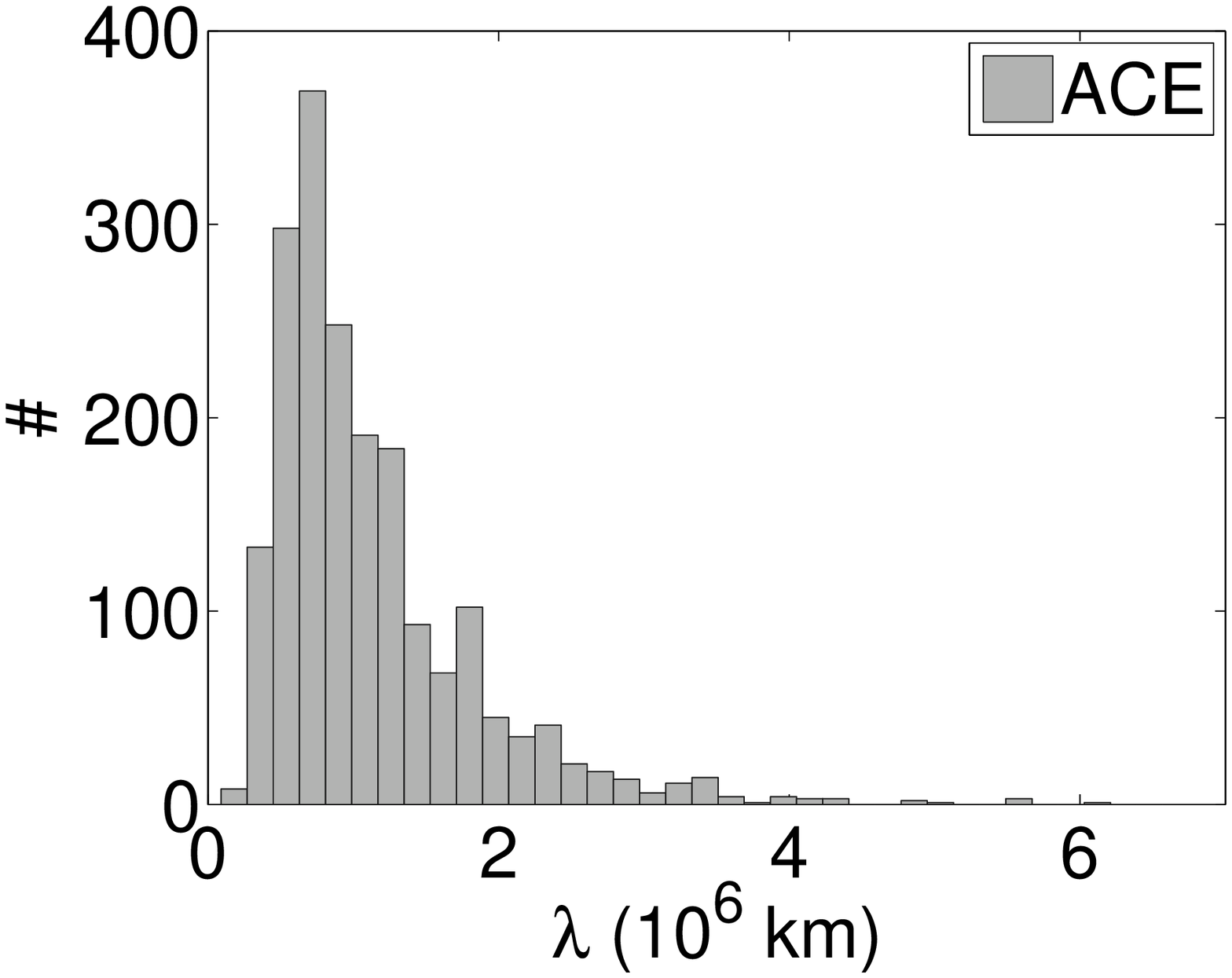}
\includegraphics[width=0.333\textwidth,height=.333\textwidth]
{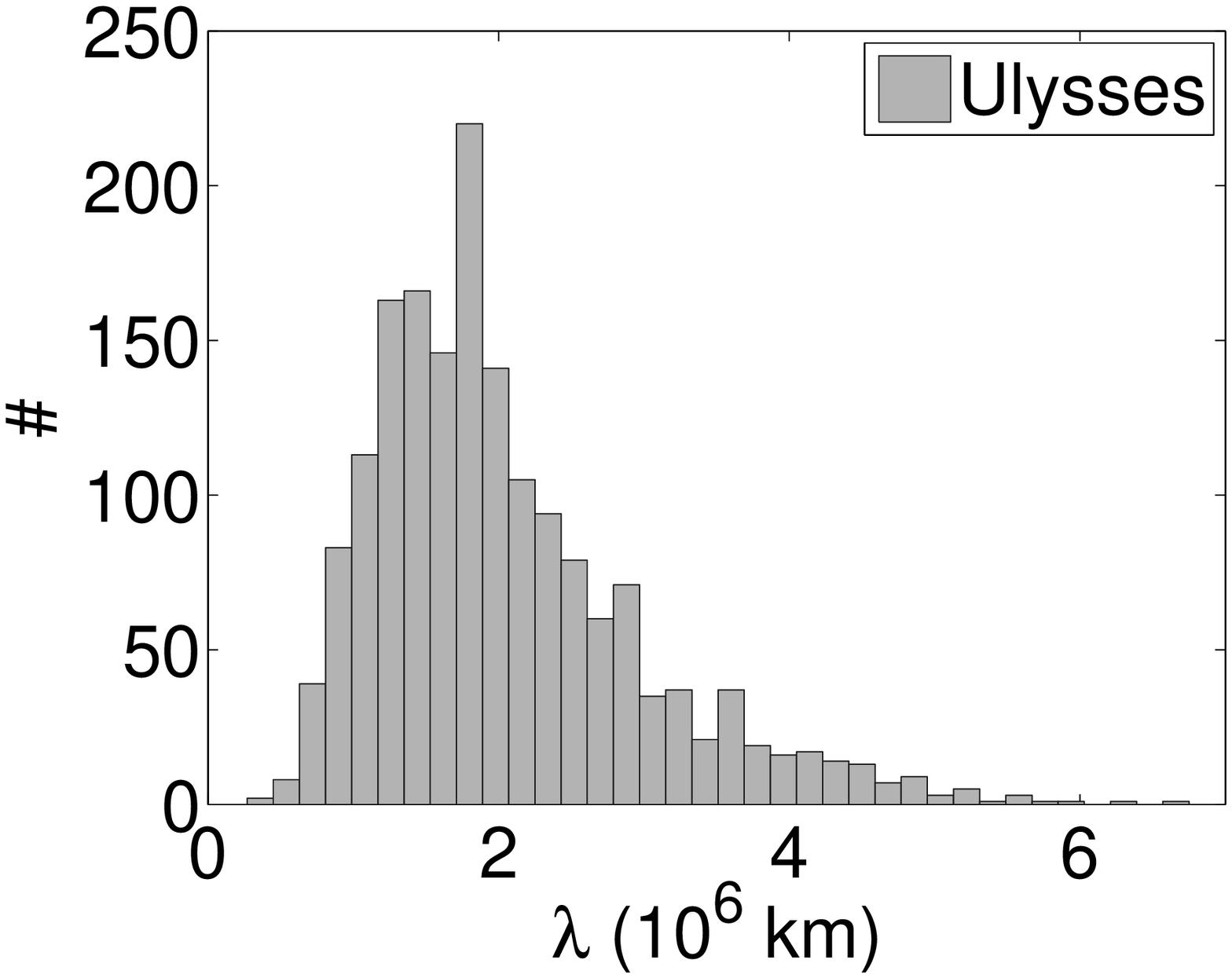}}
\vspace{-0.00\textwidth}   
\centerline{\Large \bf
\hspace{0.138\textwidth}  \color{black}{\small{(a)}}
\hspace{0.275\textwidth}  \color{black}{\small{(b)}}
\hspace{0.268\textwidth}  \color{black}{\small{(c)}}
\hfill}
\vspace{0.030\textwidth}    
\caption{Observed histograms of $\la$. Panels (a), (b), and (c)
corresponding to H1$+$H2, ACE, and {\it Ulysses} data respectively.}
\label{F_histo_lambda}
\end{figure}

Figure \ref{F_histo_lambda} presents the observed histograms of $\la$ at 
each station. Heliocentric distance increases from left to right: panels $a$, $b$, and 
$c$ corresponding to H1$+$H2, ACE, and {\it Ulysses} data, respectively. As the 
heliocentric distance increases, the bins at the right become progressively occupied.
In each case, the distribution is clearly asymmetric with a long tail on the 
right side. This long tail is evidence of non-linear interactions and multiplicative 
processes, and motivates us to explore the hypothesis of a log-normal PDF for $\la$. 

Since normal and log-normal distributions are related (see Appendix 
\ref{A_log_normal_statistics}), we perform a statistical analysis on $\ln(\la)$
by computing the different moments of the histogram of $\ln(\la)$, and then make 
use of Equations~(\ref{E_mean_lognorm}) (Appendix \ref{A_log_normal_statistics}) 
for the statistics of $\la$.

Moments of higher order than the variance will become of interest since we 
want to study how the distribution of $\ln(\la)$ deviates from Gaussianity. 
The third, fourth, and sixth central moments of a probability distribution 
function are defined as follows
\begin{equation}\label{E_moments}
 \gamma=\frac{E[(x-\mu)^3]}{\sigma^3}, \qquad
 K=\frac{E[(x-\mu)^4]}{\sigma^4},      \qquad
 M_6 =\frac{E[(x-\mu)^6]}{\sigma^6}
\end{equation}
where $E$ means expectation value, $x$ a random variable, $\mu$ its expectation value, 
and $\sigma$ its standard deviation. All odd central moments for a symmetric distribution 
are zero. Then any non-vanishing odd central moment can be taken as a measure of asymmetry 
of the distribution. Positive values of the skewness [$\gamma$] indicate that the distribution 
has a larger tail to the right of the mean value, while negative values indicate a larger tail 
to the left. The moments $K$ and $M_6$ are useful to compare with the Gaussian distribution, 
for which $K=3$ and $M_6=15$. Any distribution with a $K$ larger (smaller) than 3 will be higher 
(lower) than a Gaussian distribution with the same mean and variance. The information on how the 
tails of the distribution fall is contained in $M_6$. Values of $M_6$ higher than 15 indicate 
more slowly decreasing tails and lower values more rapidly decreasing tails than a Gaussian 
distribution.

\begin{figure}
\centerline{
\includegraphics[width=0.333\textwidth,height=.333\textwidth]
{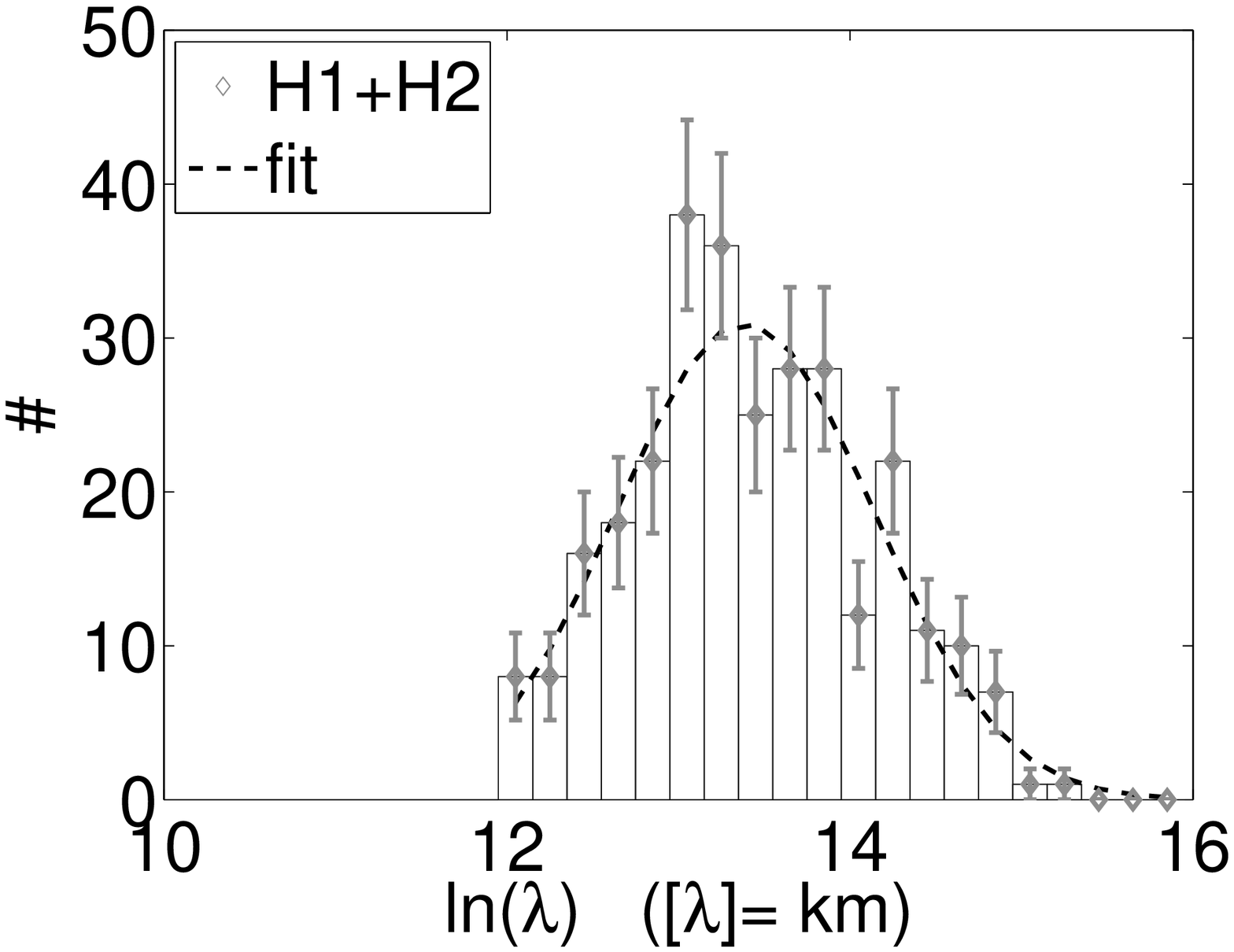}
\includegraphics[width=0.333\textwidth,height=.333\textwidth]
{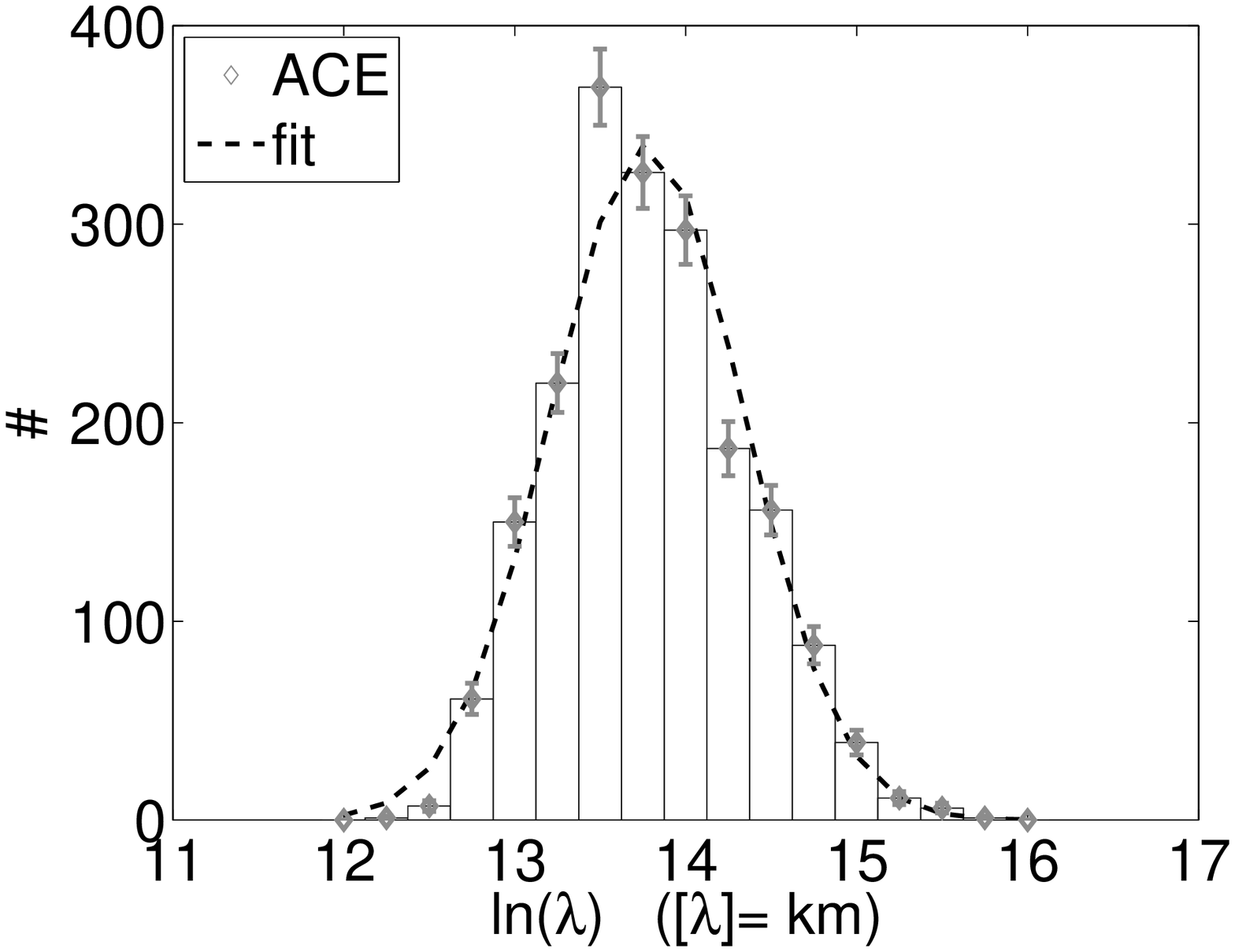}
\includegraphics[width=0.333\textwidth,height=.333\textwidth]
{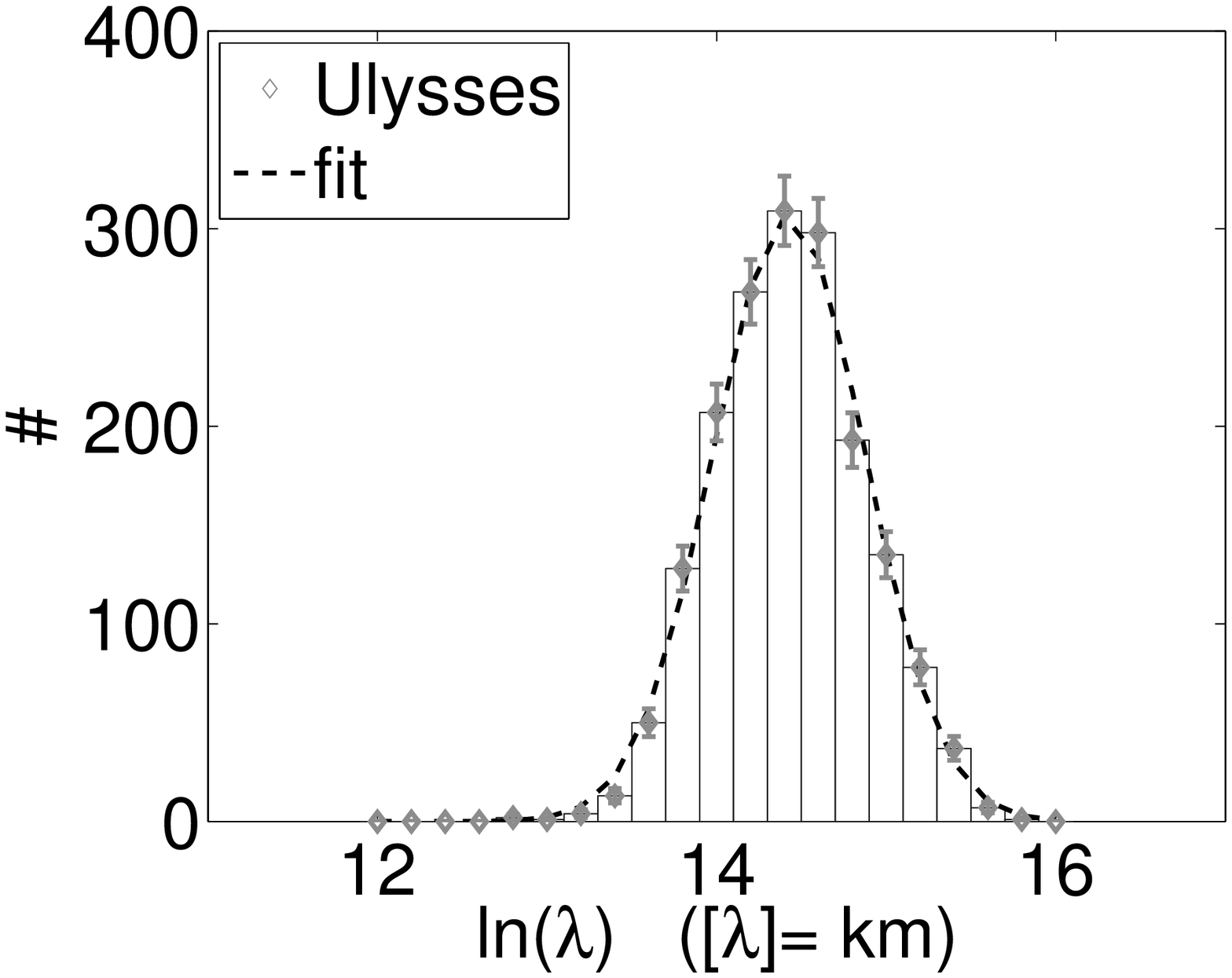}
}
\vspace{-0.00\textwidth}   
\centerline{\Large \bf     
\hspace{0.138\textwidth}  \color{black}{\small{(a)}}
\hspace{0.275\textwidth}  \color{black}{\small{(b)}}
\hspace{0.268\textwidth}  \color{black}{\small{(c)}}
\hfill}
\vspace{0.036\textwidth}    
\caption{Histograms of $\ln(\lambda)$ and non-linear least-square fit.
Panels (a), (b), and (c) correspond to H1$+$H2, ACE, and {\it Ulysses} data respectively.}
\label{F_histo_chi}
\end{figure}

Figure~\ref{F_histo_chi} presents the histograms of $\ln(\la)$, together with a non-linear 
least-squares fit to the data of a Gaussian trial function. There are empty bins at the left 
of the histogram in panel (a). This is due to the interval selection procedure followed in the 
present work. The definition of the quality factor [$F$] depends explicitly on $\la_{ii}$ 
calculated through a linear fit to $\ln(R)\approx -r/\lambda$. Due to the time cadence available, 
for correlation functions decreasing very fast it is not possible to fit a straight line since 
there is only one point available in the region of interest. These cases, which fill the bins 
to the left, were assigned with a flag and were excluded from the analysis.
  
Table~\ref{T_parameters_histos} shows the relevant statistical parameters of the distributions 
of $\la$ (see Figure~\ref{F_histo_lambda}) and $\ln(\la)$ (see Figure~\ref{F_histo_chi}).

The first panel presents the moments of the observed $\ln(\la)$ distribution, obtained 
directly from the data: mean [$\mu$] and variance [$\sigma^2$], skewness [$\gamma$], kurtosis 
[$K$] and sixth central moment [$M_6$]. We included the mean value [$m$] and median [$m^*$] of the 
$\la$ distribution (Figure~\ref{F_histo_lambda}) obtained from Equations~(\ref{E_mean_lognorm}).

The second panel presents the values of the parameters returned by the non-linear least-square 
fits to the data: mean [$\mu$] and variance [$\sigma^2$], number of degrees of freedom [$dof$], 
minimum [$\chi^2$], and mean [$m$] and median [$m^*$] of the $\la$ distribution. A comparison 
between the moments obtained from data and those obtained through the fits is shown in the third 
panel, where we report the ratios between panel 1 (P1) and panel 2 (P2) quantities. As expected, 
values obtained directly from data and values obtained from the fitting procedure are remarkably 
similar. 
 
The number of intervals [$I$] considered in each case are shown in the last row of 
Table~\ref{T_parameters_histos}.

As $D$ increases, moments (from observations and from fits) evolve to have the same values (panel 
3 of Table~\ref{T_parameters_histos}), the fourth and sixth central moments show a trend to reach 
the values expected for a Gaussian distribution, and the variance decreases. The skewness does not 
show a definite trend, but nevertheless it does not depart too much from the zero expected for 
a symmetric distribution.

\begin{table}
\caption{Statistical parameters of $\ln(\lambda[{\mathrm{km}}]$) distributions shown 
in Figure~\ref{F_histo_chi}. First panel: values from the observed PDF. Second panel: 
values from the non-linear least-squares fit. Third panel is a comparison between the 
first two. We also show in panel 4 the $Pv$ from KS goodness-of-fit test. The number 
of intervals analyzed in each case is reported in the last row.}
\label{T_parameters_histos}
\begin{tabular}{cccc}     
\toprule
\multicolumn{4}{l}{\textbf{Panel 1: From observations}}\\
\midrule 
                        & H1$+$H2 & ACE        &  {\it Ulysses} \\
\hline 
$\mu$                   & 13.42     & 13.78       & 14.43\\
$\sigma^2$              & 0.47      & 0.30        & 0.19 \\
$\gamma$                & 0.21      & 0.35        & 0.04 \\
$K$                     & 2.49      & 2.79        & 2.94 \\
$M_6$                   & 9.75      & 12.40       & 12.71\\
$m$  ($\times 10^6$~km) & 0.85      & 1.12        & 2.03 \\
$m^*$($\times 10^6$~km) & 0.67      & 0.97        & 1.85 \\
\toprule
\multicolumn{4}{l}{\textbf{Panel 2: From non-linear fit to data}}\\
\midrule 
                        &H1$+$H2    & ACE        &  {\it Ulysses} \\
\hline 
$\mu$                   & 13.39     & 13.78       & 14.43\\
$\sigma^2$              & 0.56      & 0.32        & 0.20 \\
$m$  ($\times 10^6$~km) & 0.86      & 1.13        & 2.04 \\
$m^*$($\times 10^6$~km) & 0.65      & 0.96        & 1.84 \\
$dof$			& 18	    & 15	  & 19	\\
$\chi^2$		& 17.70	    & 60.19	  & 23.5\\
\toprule 
\multicolumn{4}{l}{\textbf{Panel 3: Comparing panel 1 with panel 2}}\\
\midrule 
                        & H1$+$H2   & ACE        &  {\it Ulysses} \\
\midrule 
$\mu_{P1}/\mu_{P2}$     &1.002      &1           &1\\
$\si^2_{P1}/\si^2_{P2}$ &0.84       &0.94        &0.95\\
$m_{P1}/m_{P2}$         &0.99       &0.99        &0.995\\
$m^*_{P1}/m^*_{P2}$ 	&1.03       &1.01        &1.01\\
\toprule 
\multicolumn{4}{l}{\textbf{Panel 4: JB goodness-of-fit test}}\\
\midrule 
                        & H1$+$H2   & ACE        &  {\it Ulysses} \\
\midrule 
$Pv$                    & 0.06      &$<$0.01 & 0.71 \\
\toprule
$\#$ of intervals       & 291       & 1919        & 1731 \\
\bottomrule 
\end{tabular}
\end{table}

\section{Hypothesis Testing}\label{S_test_hip:intro}

Histograms and the different moments of the observed $\ln(\lambda)$ distribution are useful 
for the characterization of the asymptotic PDF and, although they may quantify deviations 
from Gaussianity, they are not enough to give conclusive evidence that the model, log-normal 
PDF of $\la$, is appropriate or not. We employ then a more formal method, the Jarque--Bera 
(JB) goodness-of-fit hypothesis test \cite{JB_paper,JB} to support and complement the graphical 
methods presented above in Section \ref{S_charac_histos}. 

The JB test is useful to check the normality assumption, that is, to test the hypothesis 
H$_{\mathrm 0}$ that the random variable $\ln(\lambda)$ is drawn from a normal distribution 
function with unknown mean and unknown variance, against an alternative hypothesis that $\ln(\la)$ 
does not come from a normal distribution. This test is suitable when the hypothesized 
distribution is not known and its parameters (mean and variance) have to be estimated.

The test statistic [$\xi_{\mathrm JB}$] is defined as
\begin{equation}
 \xi_{\mathrm JB}=\frac{n}{6}\left(\gamma^2 + \frac{(K-3)^2}{4}\right)
 \label{E-xi-JB}
\end{equation}
where $n$ is the number of data points, $\gamma$ is the sample skewness, and $K$ is 
the sample kurtosis; $\xi_{\mathrm JB}$ is asymptotically $\chi^2$ distributed with 
two degrees of freedom \cite{JB_paper}. Here we want to emphasize that, for the normal 
distribution the skewness and kurtosis are quantities with ​​defined values. 

We then, at each spatial station independently, implement the JB test on $\ln(\lambda)$, 
proposing that its PDF is Gaussian. We use the built-in MatLab function
and test the hypothesis at a (conventional) 0.05 significance level [$\alpha$]
(e.g., \opencite{Frodesen}).
Results are presented in the fourth panel of Table \ref{T_parameters_histos} in terms of 
the $P$-value [$Pv$] (the largest $\alpha$ that can be tolerated without 
rejecting H$_{\mathrm 0}$): values of $Pv$ larger than 
$\alpha$ indicate to accept the H$_{\mathrm 0}$, otherwise H$_{\mathrm 0}$ should
be rejected (see Appendix \ref{A_p_value}).

The values of $Pv$ obtained for H1$+$H2 and {\it Ulysses} data sets, $Pv=$0.06
and $Pv=$0.71 respectively, are strong evidence 
supporting the hypothesis thus we may accept 
that magnetic autocorrelation lengths have a log-normal distribution at a 0.06 and 0.71
significance level for H1$+$H2 and {\it Ulysses} respectively.

To understand this better, suppose that we again measure correlation lengths in the 
inner heliosphere and we present the data in a histogram. With this hypothesis being 
true, the probability of getting a histogram of $\ln(\la)$ such as, or worse than, 
the one presented here (Figure \ref{F_histo_chi}(a)) is given by $Pv$, that is a 
probability of 6$\,\%$. For the outer heliosphere, the result is even better, since 
the probability of obtaining a histogram like, or worse than, the one in Figure 
\ref{F_histo_chi}(c) is higher, meaning that the one reported here is one of the best.

Regarding ACE data, although results are qualitatively good, quantitative 
evidence, stated through $Pv$, is not sufficient to conclude that the 
$\la$-distribution is log-normal. We revisit this issue in the next Section. 

\section{Low and High Proton $\be_{\mathrm p}$}

In the last section the data employed in the analysis was 
selected for latitude and data quality, but not selected 
according to values of plasma parameters. Here we include 
an analysis based on a familiar parameter, the proton beta 
[$\be_{\mathrm p}$].  
In the SW dynamics, high or low values of $\be_{\mathrm p}$  
(proton kinetic pressure/magnetic pressure) indicate the 
dominant role of the gas or the magnetic field, respectively. 

In low and high $\be_{\mathrm p}$ regimes different wave modes can be triggered.
There are also relations between $\be_{\mathrm p}$ and, for example, temperature 
anisotropy which introduce constraints to the system and determine the properties 
of the instabilities in space plasma conditions ({\it e.g.}, \opencite{Bale2009}). 
Thus, $\be_{\mathrm p}$ is a parameter that play a key role in the regulation of 
waves propagation and the triggering of instabilities in the SW.
 
In this section then we further investigate the statistics of 
correlation lengths when subdividing the sample into two 
groups with high and low values of $\be_{\mathrm p}$.

As limiting $\be_{\mathrm p}$ values we choose $\be_{\mathrm p}=$\,0.4 to define a 
low-$\be_{\mathrm p}$ regime, and $\be_{\mathrm p}=$\,0.7 for the high-$\be_{\mathrm p}$ 
regime, in order to allow a clear division between the sets while keeping a 
statistically significant amount of data.

\begin{figure}
\centerline{
\includegraphics[width=0.453\textwidth,height=.333\textwidth]{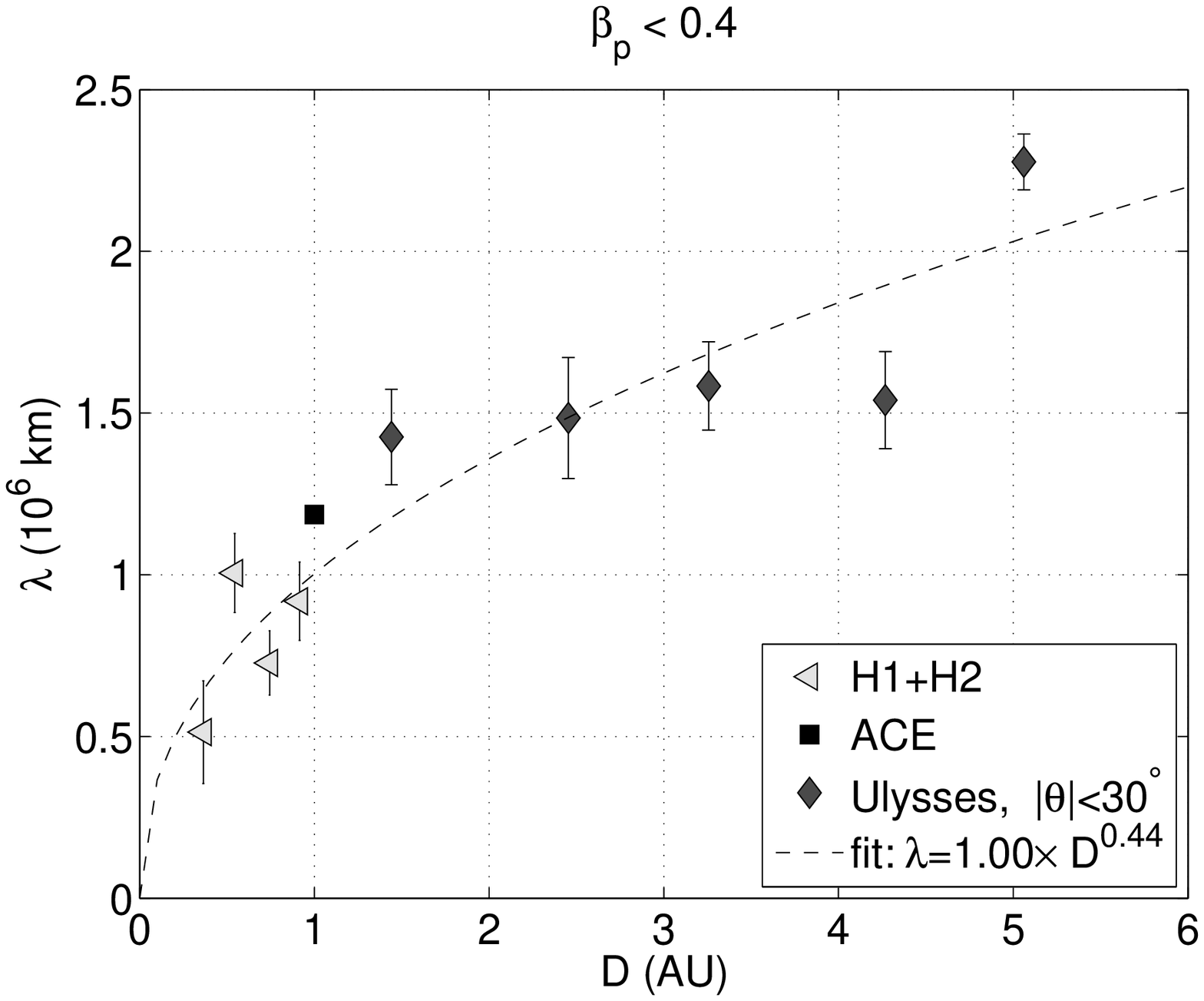}
\includegraphics[width=0.453\textwidth,height=.333\textwidth]{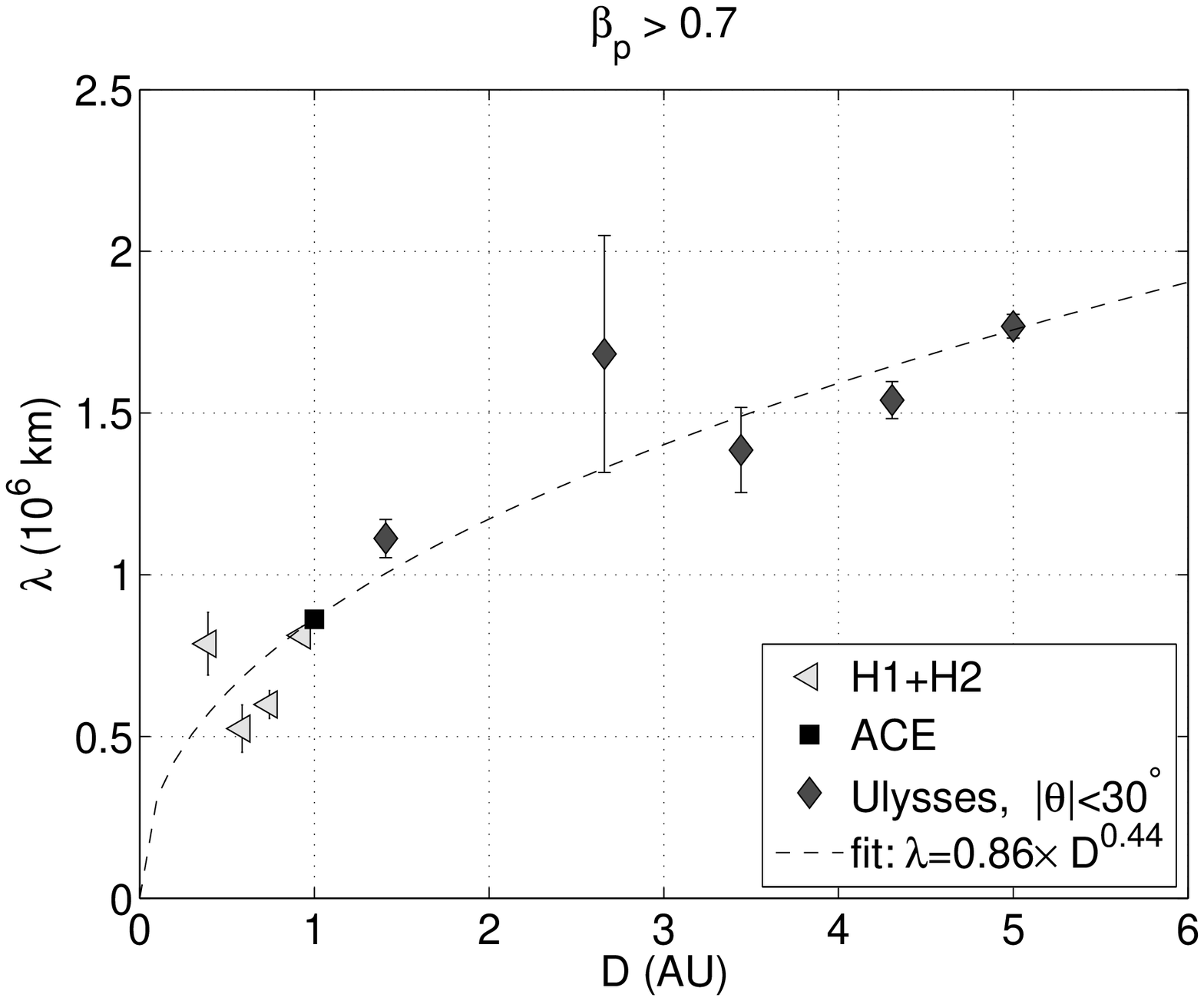}
}
\vspace{-0.00\textwidth}   
\centerline{\Large \bf     
\hfill}
\vspace{0.030\textwidth}    
\caption{Plot of observed $\la$ versus heliocentric distance for 
low-$\be_{\mathrm p}$ and high-$\be_{\mathrm p}$ regimes.}
\label{F_dist_beta_evol}
\end{figure}

\begin{figure}
\centerline{
\includegraphics[width=0.453\textwidth,height=.333\textwidth]{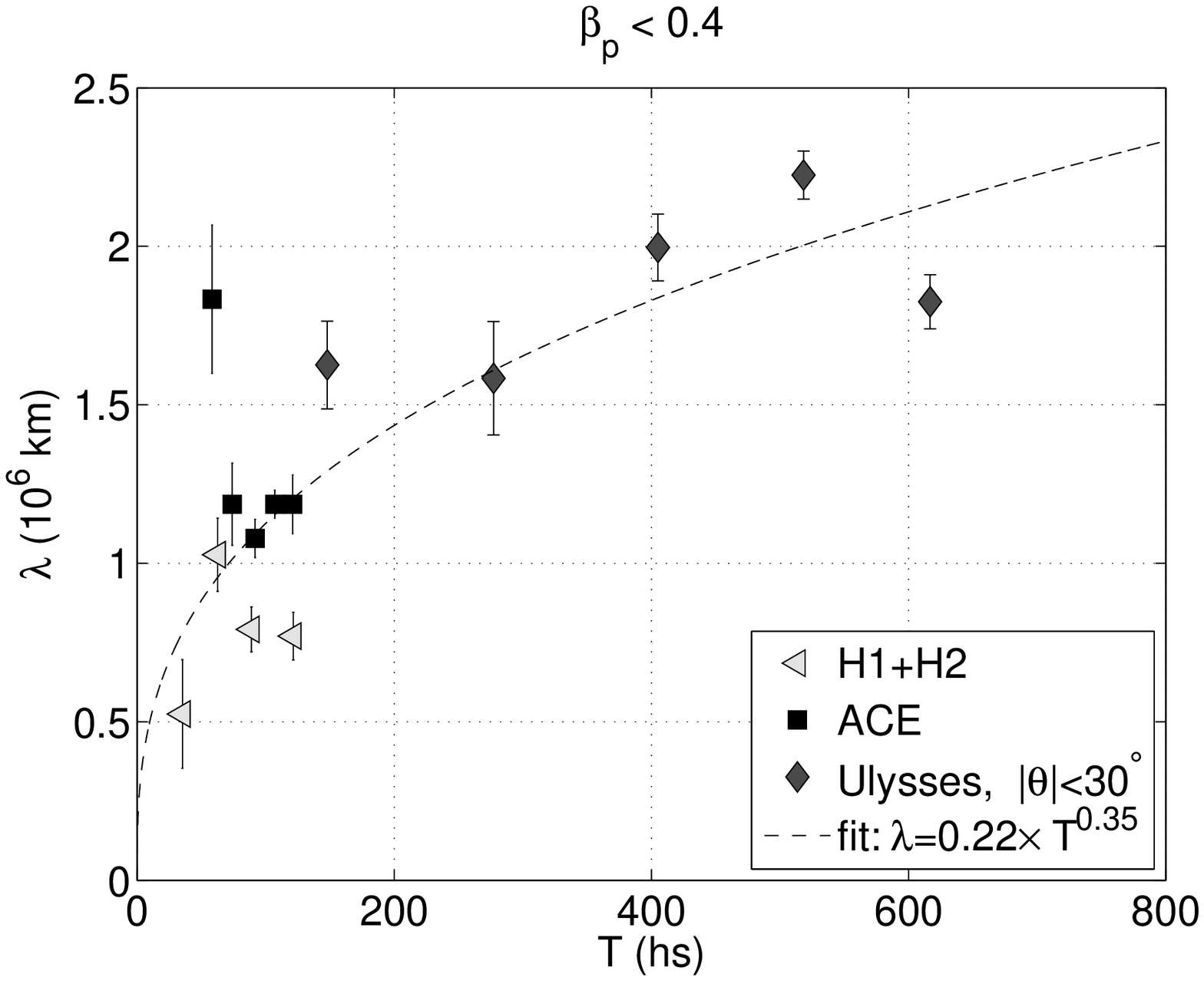}
\includegraphics[width=0.453\textwidth,height=.333\textwidth]{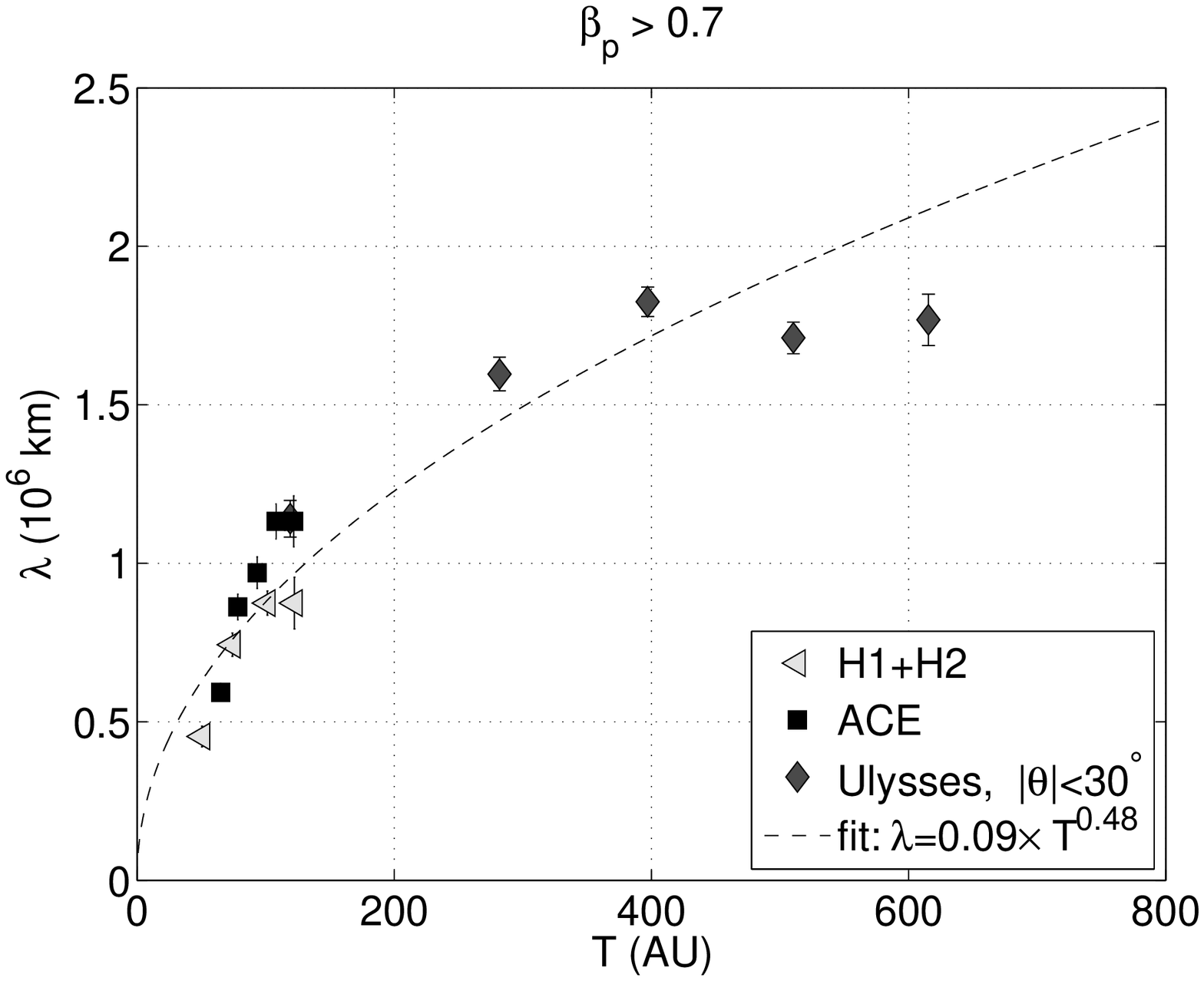}
}
\vspace{-0.00\textwidth} 
\centerline{\Large \bf   
\hfill}
\vspace{0.030\textwidth}    
\caption{Plot of observed $\la$ versus solar-wind age for 
low-$\be_{\mathrm p}$ and high-$\be_{\mathrm p}$ regimes.}
\label{F_age_beta_evol}
\end{figure}

Figures \ref{F_dist_beta_evol} and \ref{F_age_beta_evol} show
how correlation lengths grow with heliocentric distance and age in both
regimes: $\be_{\mathrm p}<$\,0.4 and $\be_{\mathrm p}>$\,0.7. We group the observations 
into bins of heliocentric distance of different widths: 
$\Delta D =$\,0.17 AU for H1$+$H2 data and $\Delta D =$\,0.85 AU for 
{\it Ulysses} data; and into bins in turbulence age of widths $\Delta T =$\,28 hours,
16 hours, 125 hours for H1$+$H2, ACE, and {\it Ulysses} respectively. 
A least-squares fit to the data yields a power-law increase in each
case; see inset in Figures \ref{F_dist_beta_evol} and 
\ref{F_age_beta_evol}.
It seems that observations of $\la$ are better ordered for $\be_{\mathrm p}>$\,0.7 
and with $T$.

\begin{figure}
\centerline{
\includegraphics[width=0.333\textwidth,height=.333\textwidth]
{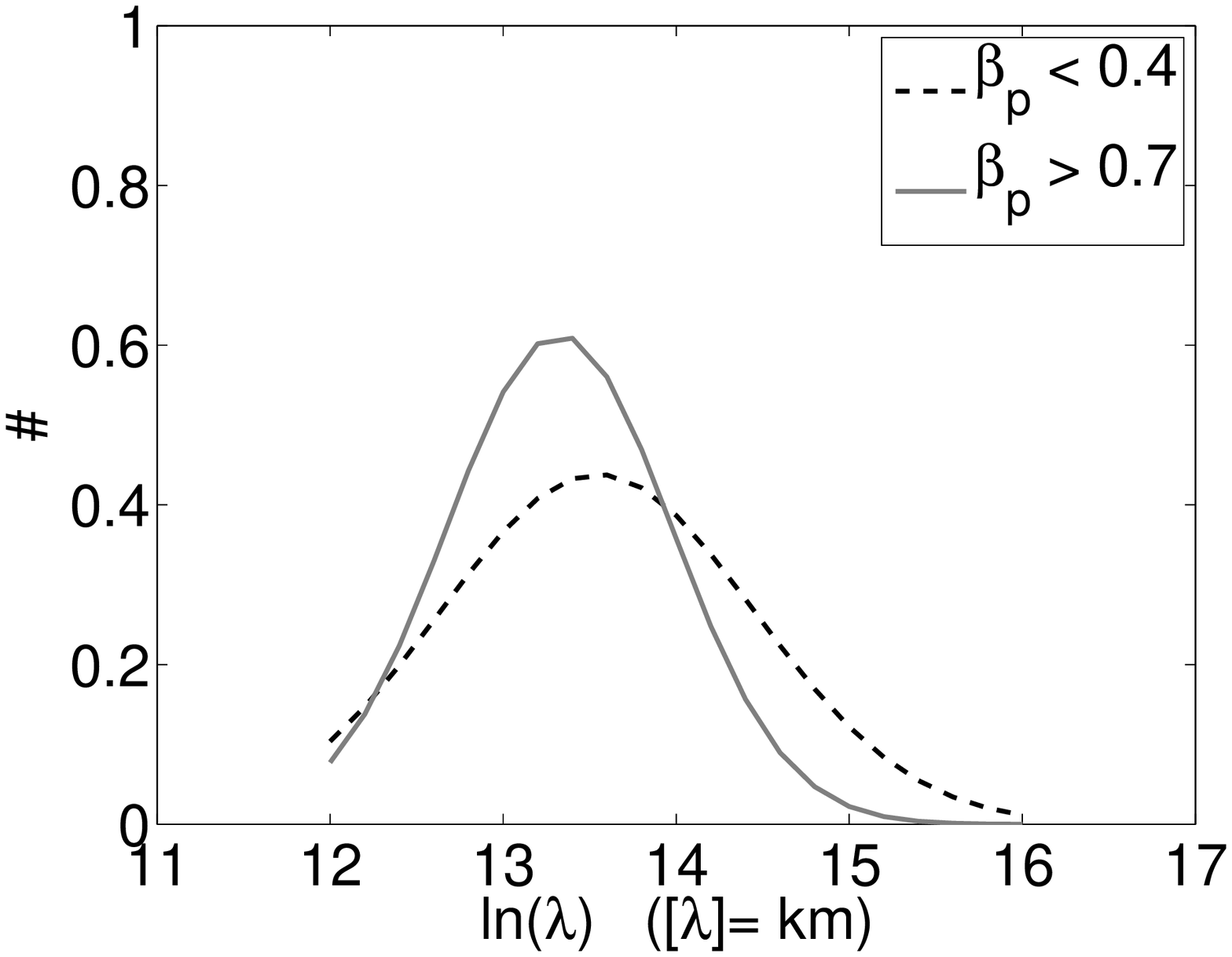}
\includegraphics[width=0.333\textwidth,height=.333\textwidth]
{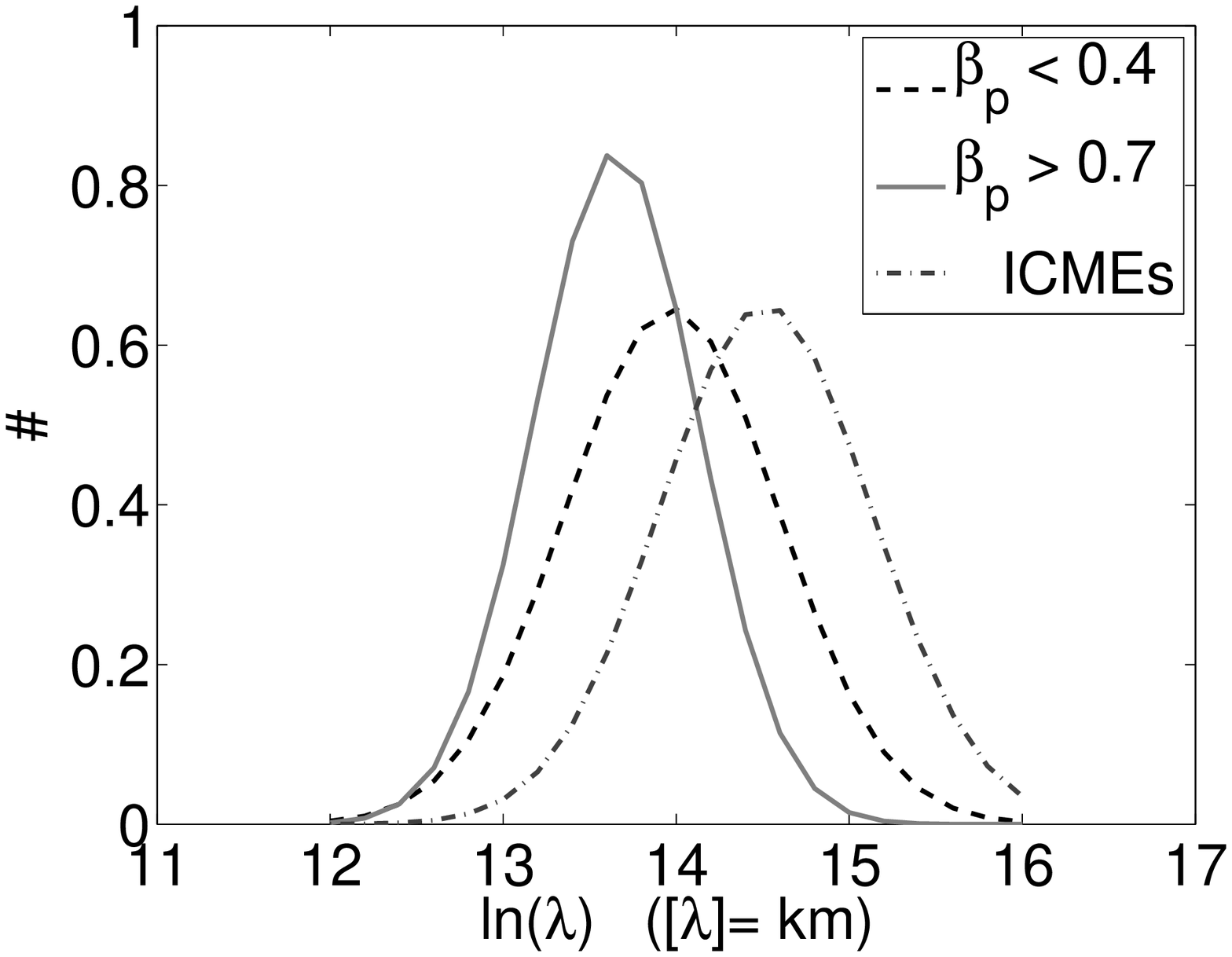}
\includegraphics[width=0.333\textwidth,height=.333\textwidth]
{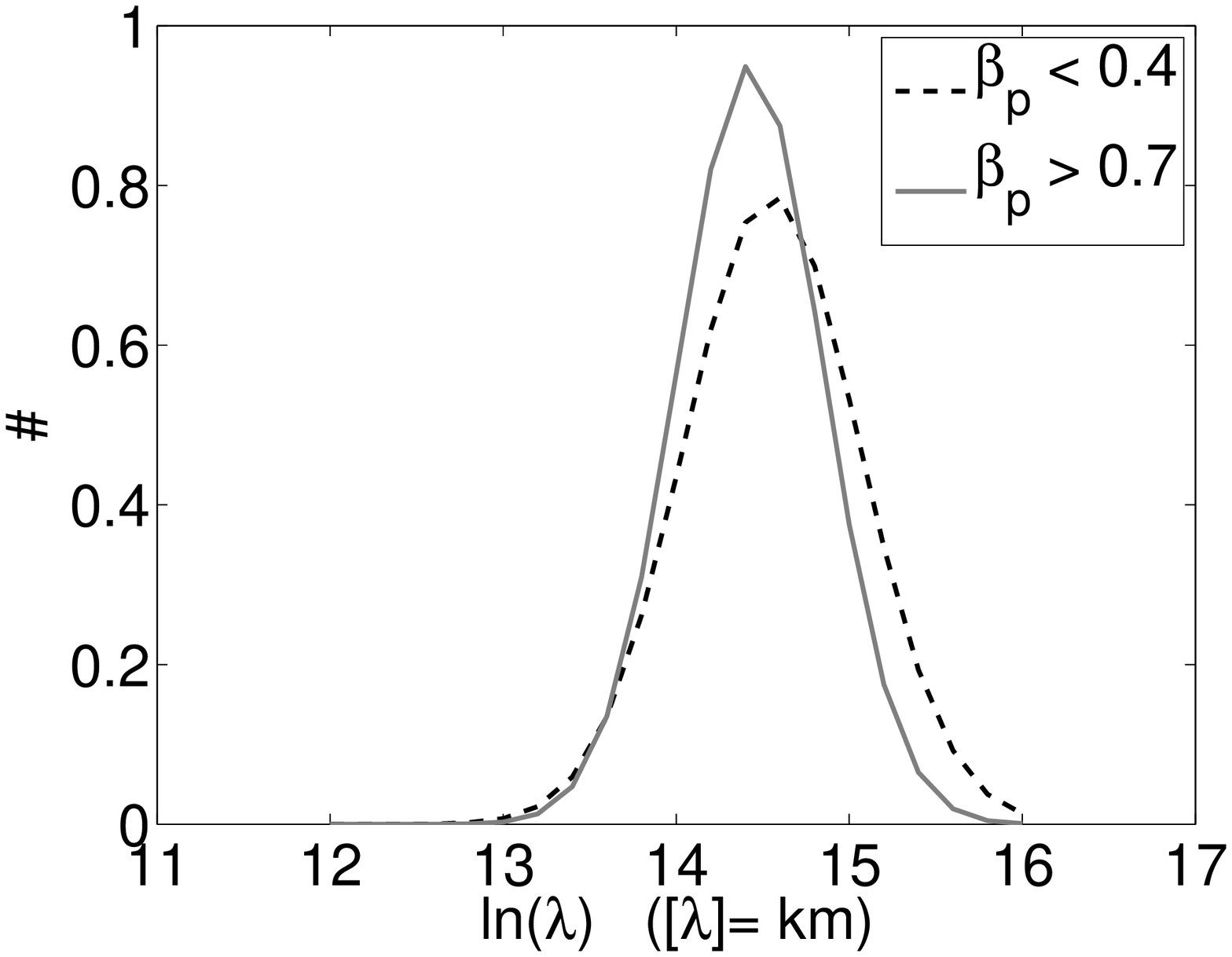}
}
\vspace{-0.00\textwidth}   
\centerline{\Large \bf     
\hspace{0.138\textwidth}\vspace{-0.02\textwidth}  \color{black}{\small{(a)}}
\hspace{0.275\textwidth}  \color{black}{\small{(b)}}
\hspace{0.280\textwidth}  \color{black}{\small{(c)}}
\hfill}
\vspace{0.030\textwidth}    
\caption{Non-linear least-square fit for high-$\be$ (solid line) and low-$\be$ 
(dashed line) samples. Panels (a), (b) and (c) correspond to H1$+$H2, ACE and 
{\it Ulysses} data respectively. 
{The dashdot line in panel (b) corresponds to the non-linear 
least-square fit for ICMEs (see Section~\ref{s:transient}).}}
\label{F_chi_beta}
\end{figure}

In Figure \ref{F_chi_beta} we present the non-linear least-square fit
to the distributions of $\ln(\la)$ for the low-$\be_{\mathrm p}$ and 
high-$\be_{\mathrm p}$ groups together. 
Here, each of the Gaussian fits is normalized by the total area, and 
therefore the value of the ordinate represents a probability density. 
Again, heliocentric distance increases from left to right, each panel corresponding
to H1$+$H2, ACE, and {\it Ulysses} observations respectively.
The $\ln(\la)$ distribution for the low-$\be_{\mathrm p}$ plasma is wider and is 
displaced to the right when compared to the high-$\be_{\mathrm p}$ plasma samples.
Table \ref{T_tabla_low_high_beta} shows the relevant statistical parameters
of the distributions shown in Figure \ref{F_chi_beta}. It has the same 
structure as Table \ref{T_parameters_histos}: the first panel presents the moments 
directly obtained from the data while the fitting parameters are listed in the 
second panel. The third panel compares results reported in panel 1 and panel 2.

For both regimes, we again find that correlation lengths grow with increasing heliocentric 
distance. Moreover, we find that both populations grow approximately at the same rate, with 
the mean correlation length in the low-$\be_{\mathrm p}$ plasma being 1.5 times higher than 
the mean $\la$ in high-$\be_{\mathrm p}$ plasma at {\it Helios} heliocentric distances, and 
diminishing to a ratio of 1.2 in the outer heliosphere.

We also perform a JB test on the three low-$\be_{\mathrm p}$ and three high-$\be_{\mathrm p}$ 
groups of correlation lengths to test if, when separating the observations into these two regimes, 
the distributions of $\la$ are consistent with a log-normal distribution. The test is performed on 
the $\ln(\la)$ distributions, and the $Pv$ values obtained are listed in the fourth panel of Table 
\ref{T_tabla_low_high_beta}. The values of $Pv$, 0.06, 0.06 and 0.10 in the low-$\be$ regime, and 
0.14, 0.08, 0.36 in the high-$\be_{\mathrm p}$ regime for H1$+$H2, ACE, and {\it Ulysses} respectively
(all of them larger than $\alpha$),
indicate that the hypothesis \textquotedblleft correlation lengths are log-normally distributed in 
low-$\be_{\mathrm p}$ SW and in high-$\be_{\mathrm p}$ SW\textquotedblright~can be accepted in all
six cases at the respective $Pv$ significance.
Note that the results for ACE  are greatly improved relative to the last Section which did not sort 
the data by $\be_{\mathrm p}$. 
Here we see that low-$\beta_p$ and high-$\beta_p$ populations observed
by ACE spacecraft are slightly separated, more than for {\it Helios} and {\it Ulysses} 
data, being the JB test is sensitive to this separation.

\begin{table}
\caption{Characterization of $\lambda$ distribution for 
low-$\beta_p$, high-$\beta_p$ regimes and ICMEs observed only by ACE}
\label{T_tabla_low_high_beta}
\begin{tabular}{lccccccc}
\toprule
\multicolumn{8}{l}{\textbf{Panel 1: From observations}}\\
\midrule 
                     &\multicolumn{3}{c}{$\be_{\mathrm p}<$0.4}&\multicolumn{3}{c}{$\be_{\mathrm p}>$0.7}&{ICMEs}\\
	             &H1$+$H2 &ACE  &{\it Ulysses} &H1$+$H2&ACE  &{\it Ulysses}& {ACE}\\ 
\toprule
$\mu$      		  &13.57  &13.98&14.55   &13.35    &13.68&14.42& {14.52}\\
$\sigma^2$ 		  &0.61   &0.35 &0.24    &0.32     &0.20 &0.17 & {0.53 }\\
$\gamma$   		  &0.13   &0.07 &-0.17   &0.41     &0.18 &0.10 & {-0.16}\\
$K$ 			  &1.96   &2.58 &2.65    &2.61     &2.62 &3.11 & {3.08}\\
$M_6$ 			  &4.88   &10.10&11.57   &8.88     &10.33&15.27& {14.00}\\
$m$~($\times 10^{6}$ km)   &1.06   &1.40 &2.35    &0.74     &0.97 &1.99 & {2.33}\\
$m^*$~($\times 10^{6}$ km) &0.78   &1.18 &2.08    &0.63     &0.87 &1.83 & {2.03}\\
\toprule
\multicolumn{8}{l}{\textbf{Panel 2: From non-linear fit to data}}\\
\midrule
                      &\multicolumn{3}{c}{$\be_{\mathrm p}<$0.4}&\multicolumn{3}{c}{$\be_{\mathrm p}>$0.7}&{ICMEs}\\
	              &H1$+$H2 &ACE  &{\it Ulysses} &H1$+$H2&ACE  &{\it Ulysses}& {ACE }\\ 
\midrule
$\mu$      		  	&13.54  &13.97&14.55   &13.32    &13.65&14.43&{14.50}\\
$\si^2$				&0.83   &0.38 &0.26    &0.42     &0.22 &0.18 &{0.59} \\
$m$~($\times 10^{6}$ km)   	&1.15   &1.41 &2.37    &0.75     &0.95 &2.02 &{2.36}\\
$m^*$~($\times 10^{6}$ km)	&0.78   &1.17 &2.08    &0.61     &0.85 &1.85 &{1.98}\\
$dof$			 	&19     &19   &19      &19       &19   &19   &{17}\\
$\chi^2$			&18     &24   &17.6    &19       &19   &20.6 &{21}\\ 
\toprule 
\multicolumn{8}{l}{\textbf{Panel 3: Comparison between panel 1 (P1) and panel 2 (P2)}}\\
\midrule 
                      &\multicolumn{3}{c}{$\be_{\mathrm p}<$0.4}&\multicolumn{3}{c}{$\be_{\mathrm p}>$0.7}&{ICMEs}\\
                      & H1$+$H2 & ACE      & {\it Ulysses}& H1$+$H2 & ACE &  {\it Ulysses}&{ACE } \\
\midrule
$\mu_{P1}/\mu_{P2}$       &1.002  &1.001&1       &1.002    &1.002&0.999&{1.00}\\    
$\si^2_{P1}/\si^2_{P2}$   &0.73   &0.92 &0.92    &0.76     &0.91 &0.94 &{0.90} \\
$m_{P1}/m_{P2}$           &0.923  &0.99 &0.99    &0.99     &1.02 &0.99 &{0.99}\\
$m^{*}_{P1}/m^{*}_{P2}$   &1      &1.01 &1       &1.03     &1.02 &0.99 &{1.03}\\
\toprule 
\multicolumn{8}{l}{\textbf{Panel 4: JB goodness-of-fit test}}\\
\midrule 
                          &\multicolumn{3}{c}{$\be_{\mathrm p}<$0.4}&\multicolumn{3}{c}{$\be_{\mathrm p}>$0.7}&{ICMEs}\\
                          & H1$+$H2 & ACE      & {\it Ulysses}& H1$+$H2 & ACE &  {\it Ulysses}&{ACE} \\
\midrule
$Pv$			 &0.06   &0.06 &0.10    &0.14    &0.08 &0.36&{0.29 }\\ 
\toprule
$\#$ of intervals         &101    &675  &409     &85      &405  &872&{119}\\
\bottomrule
\end{tabular}
\end{table}

\section{Interplanetary Coronal Mass Ejections}
\label{s:transient}

A typical example in the SW of low $\be_p$ are magnetic clouds (MCs), transient regions 
observed having a smooth rotation of the magnetic-field direction, a magnetic field 
strength higher than average SW, low proton temperature and thus, low $\be_{\mathrm p}$ 
compared to the ambient SW. Typical values for $\be_{\mathrm p}$ at 1 AU for ambient SW 
are $\approx$ 0.6 \cite{Mullan2006}, while in clouds $\be_{\mathrm p}$ is typically around
0.1 \cite{lepping2003}. Also, they have lower turbulence levels and different turbulent 
properties ({\it e.g.}, \opencite{Dasso2003JGRA}; \opencite{Bill-Taylor-scale-2008ApJ}).

MCs are a subset of ICMEs which are also characterized by abnormally low proton temperatures, 
bidirectional streaming of suprathermal electrons and energetic ions, enhanced helium abundance,
strong magnetic fields with smooth field rotations, {\it etc.} ({\it e.g.}, 
\opencite{Neugebauer1997}, and references therein).

To study the distribution of correlation lengths in this transient component of the SW, we 
employ the defining criterion by \inlinecite{RC1995JGR} and retain those intervals showing 
an observed temperature lower than one-half of the expected temperature for usual SW \cite{Lopez_y_Freeman_1986JGR,Pascal09SolPhys}. 
Since the analysis of this transient component reduces significantly the amount of available data, 
we use only ACE data and broaden the range of $F$ to $($5$^{\mathrm{th}}$-95$^{\mathrm{th}})$ 
percentiles to increase the amount of intervals to be analyzed. 
 
The non-linear least-square fit to the distribution of $\ln(\la)$ in this transient
regime is shown in panel (b) of Figure \ref{F_chi_beta} in dashdot line. 
The Gaussian fit is also normalized by the total area to facilitate comparison. 
The $\ln(\la)$ distribution for the transient regime is also wider than the 
high-$\be_{\mathrm p}$ plasma samples and is even more displaced to the right 
when compared to the low-$\be_{\mathrm p}$.
JB test return a P$-$value equal to 0.29 giving good evidence favoring
the hypothesis \textquotedblleft correlation lengths in ICMEs follow
a log-normal PDF\textquotedblright. The last column in Table 
\ref{T_tabla_low_high_beta} shows the statistical parameters
for this distribution. 

Correlation lengths in ICMEs at 1 AU have a distribution similar to that of
low-$\be_{\mathrm p}$ plasma but with a greater mean (see Table \ref{T_tabla_low_high_beta}).

\section{Summary, Discussion, and Conclusions}\label{S_Conclusion}  

The spatial scales associated with the correlation length [$\la$] are related to 
the breakpoint in the spectrum, which separates the inertial range from the injection 
(low-frequency) range associated with large-scale structures in the SW ({\it e.g.},
presence of velocity shear). During the expansion of the wind, this breakpoint moves 
to the lower-frequency part of the spectrum \cite{TUetal1984,BrunoRev2005}.

In this work, we have analyzed {\it Helios} 1 and 2, ACE, and {\it Ulysses} magnetic 
observations, restricted to the Ecliptic plane for different heliocentric distances [$D$]. 
From these observations, we characterized the distributions of $\lambda$ in the solar 
wind, 
in low and high proton $\be$ SW regimes, and in ICMEs at 1 AU.
We quantitatively investigated the hypothesis that the PDF of $\lambda$ is log-normal.

In particular, we fitted the two free parameters of a normal distribution to the 
observed PDF of $\ln(\la$). 
Qualitatively and with respect to the fitted parameters, all of the samples
appear to be reasonably well described by a log-normal distribution.
Then we applied the Jarque-Bera goodness-of-fit test in order to quantify 
departures from log-normality of the PDFs.
We find, in the case of H1$+$H2 and {\it Ulysses} data, clear evidence 
(\textit{i.e.}, $Pv>\alpha=0.05$)
in favor of concluding that $\la$ is log-normally distributed. 
On the other hand, evidence is not so conclusive regarding ACE data: such a low 
$Pv$ indicates that we should reject the hypothesis.
 
We also studied the distribution of correlation lengths in low-$\be_{\mathrm p}$ 
and high-$\be_{\mathrm p}$ SW regimes and estimated moments of the distribution.
In each case, moments of $\ln(\la)$ evolve towards what is expected for a Gaussian 
PDF. Evaluation of the hypothesis of a normal distribution for $\ln(\la)$ by means 
of the JB test yield $Pv>\alpha=0.05$ in all cases. 
We conclude that the distribution of 
magnetic-correlation lengths can be regarded as log-normal
when considering individually the low-$\be_{\mathrm p}$ and high-$\be_{\mathrm p}$ 
solar-wind regimes. Evidently the conclusion is now equally strong for all three 
spacecraft; the identification of a log-normal distribution in the ACE analysis is 
much more conclusive when the data was sorted by proton $\be$.

Furthermore, the $\la$-distribution for the high-$\be_{\mathrm p}$ plasma is narrower 
and displaced to the left with respect to the low-$\be_{\mathrm p}$ case. While in 
the former regime the fluctuating amplitude is larger than in the latter, correlation 
lengths take smaller values in the former (high-$\be_{\mathrm p}$ sample), contrary 
to what is expected from MHD turbulence theories such as the K\'arm\'an and Howarth HD 
approach \cite{Karman-Howarth-38}. We interpret this behavior of $\la$ as a consequence 
of the different initial conditions of the magnetic-field fluctuations at the solar 
corona for the two kinds of solar wind, with the initial $\la$ in high-$\be_{\mathrm p}$ 
smaller than in the low-$\be_{\mathrm p}$ SW.

Besides its stationary component, the SW has several transient components, 
of which ICMEs are an example. We separate this transient component 
(only at 1 AU) from the usual SW retaining intervals with an observed temperature 
lower than one-half of the expected temperature for usual SW \cite{RC1995JGR}.
The distribution of $\la$ is similar to but displaced to the right with respect
to the low-$\be_{\mathrm p}$ case. The JB test yields in this case the largest $Pv$
so the hypothesis of a log-normal PDF for $\la$ can be again accepted.

The PDF of $\lambda$ evolves with the distance to the Sun. For larger heliocentric 
distances we found a narrower distribution (a decreasing $\sigma^2$ with $D$), and 
nearer to a log-normal distribution of $\lambda$. From Table 1 is possible to see that 
for increasing heliocentric distance, the moments of the PDF of $\ln(\lambda$) [$\gamma$, 
$K$, and $M_6$] tend progressively to those values expected for a normal distribution. 
This result is consistent with multiplicative processes involving $\lambda$ occurring 
in the solar wind, and a consequent relaxation to a log-normal PDF.
We confirmed that $\lambda$ increases with the heliocentric distance [$D$] and with the 
nominal SW aging [$T=D/V_{SW}$], and found that 
$\lambda(D)= 0.89 (D/1\,AU)^{0.43}\times10^6$\,km and 
$\lambda(T)= 0.11 (T/1\,\mathrm{hour})^{0.47}\times10^6$\,km, 
for the ranges [0.3\,-\,5.3]AU and [30\,-\,670] hours, respectively. We find this overall 
behavior also in the low-$\be_{\mathrm p}$ and high-$\be_{\mathrm p}$ regimes:
$\la$ grows with $D$ and $T$ in both cases. 

In the near-Ecliptic structured solar wind, fluctuations of the magnetic field 
are present over a large range of spatial and temporal scales. These multiscale 
structures partially originate at the Sun and evolve due to the local turbulent 
dynamics in the solar wind. In this context we infer that near the Sun, before 
the Alfv\'enic critical point, $\la$ follows a log-normal probability distribution 
function in both high-$\be_{\mathrm p}$ and low-$\be_{\mathrm p}$ solar wind, whose 
parameters continue to evolve due to the solar-wind turbulent dynamics. The 
distribution remains approximately log-normal, and evolves more precisely towards 
this form due to multiplicative processes in the turbulent solar wind.

\begin{acks}
MER is a fellow from CONICET. SD is a member of the Carrera del Investigador 
Cient\'{\i}fico, CONICET. MER and SD acknowledge partial support by Argentinean 
grants UBACyT 20020120100220 (UBA) and PIP 11220090100825/10 (CONICET). 
WHM acknowledge partial support by NFS Shine AGS 1156094, Solar Terrestrial 
Program AGS 1063439, and the Solar Probe Plus ISIS Project.
We thank E. Marsch for providing {\it Helios} data.
\end{acks}
\appendix   

\section{Log-normal PDFs} \label{A_log_normal_statistics} 
   
A probability distribution function (PDF) of variable $x$ is said to be log-normal if:
\begin{equation}\label{E_def_lognorm}
y=f_L(x|\mu,\sigma)=\frac{1}{x \sigma \sqrt{2\pi}}\exp{\biggl[-\frac{(\ln(x)-\mu)^2}{2\sigma^2}\biggr]}
\end{equation}
The Gaussian (or normal) distribution and the log-normal distribution 
are related. Let $Y$ be a random variable log-normally distributed with
parameters $\mu$ and $\sigma$, then $x=\ln(Y)$ will be a random variable
normally distributed with mean $\mu$ and standard deviation $\sigma$.
The mean [$m$], median [$m^*$], and variance [$var$] of $Y$ are functions of 
the parameters $\mu$ and $\sigma$ \cite{Mood} given by:
\begin{equation}\label{E_mean_lognorm}
 m=\exp{\biggl(\mu + \frac{\sigma^2}{2}}\biggr), \quad 
 m^*=\exp{(\mu)}, \quad
 var=[\exp(\sigma^2) - 1]\exp(2\mu + \sigma^2)
\end{equation}
and $\mu$ and $\sigma$ are the mean and standard deviation of 
$x=\ln(Y)$ (normally distributed). 

Log-normal forms are a possible fit when the distribution of 
a positive-definite random variable has a low mean, large variance,
and is asymmetric with long tail to high values, larger than 
the mean value \cite{Limpert01}.
Log-normal PDFs are usually encountered when the observable variable results 
from a large number of independent processes operating simultaneously.
The long tail exhibits non-linear interactions and multiplicative 
processes, thus making the log-normal PDF suitable for the description of
highly variable observations.

\section{Goodness-of-fit Hypothesis Test and P$-$value} \label{A_p_value} 

When fitting a statistical model to observed data, one may want to know
how well the model actually reflects the observations. That is, how close 
are the observed values to those which would be expected under the fitted model?
There are statistical hypothesis tests that addresses this issue.

Any hypothesis test uses a statistic $\xi$, namely a certain quantity calculated from 
the data whose probability distribution function [$f$] is known, assuming that the 
hypothesis to be tested is true. In the particular case of the Jarque--Bera \cite{JB_paper} 
test, $\xi_{\mathrm JB}$ was already introduced in Equation (\ref{E-xi-JB}): 
$\xi_{\mathrm JB}=\frac{n}{6}\left(\gamma^2 + \frac{(K-3)^2}{4}\right)$. In order 
to accept or reject the hypothesis we need a decision rule. If the computed 
$\xi_{\mathrm{obs}}$ is larger than a critical value $\xi_{\mathrm{critical}}$ 
(chosen \textit{a priori}), then the observed and expected values are not close 
enough and the model is a poor fit to the data. 

One can state the decision rule in probabilistic terms. The probability of rejecting 
a true hypothesis is the significance $\alpha$
\begin{equation}\label{E_alpha}
 \alpha=\int_{\xi_{{\mathrm critical}}}^{\infty} f(\xi){\mathrm d}\xi
\end{equation}
Stated in this way, $\alpha$ determines the critical value $\xi_{\mathrm{critical}}$ 
of the statistic in use \cite{Frodesen}. Then, if after conducting the test, our $\xi$ 
yields an observed value $\xi_{\mathrm{obs}}$ greater than $\xi_{\mathrm{critical}}$ 
({\it i.e.} $\xi_{\mathrm{obs}} < \xi_{\mathrm{critical}}$), we should reject our hypothesis.
Increasing $\alpha$ will increase the probability of incorrectly rejecting the hypothesis 
when it is true.

However, it is even more convenient to calculate the $P$-value $Pv$, defined, 
assuming the hypothesis to be true, as
\begin{equation}\label{E_pvalue}
 Pv=\int_{\xi_{obs}}^{\infty} f(\xi){\mathrm d}\xi
\end{equation}
The $P$-value is the highest value of $\alpha$ that we can obtain from the 
test such that we do not reject the null hypothesis \cite{Frodesen}. 
Statistically speaking, the $P$-value is the probability of obtaining 
a result as extreme as, or more extreme than, the result actually obtained 
when the null hypothesis is true. 
The $P$-value (obtained for $\xi_{\mathrm{obs}}$) can be understood as follows: 
suppose that we perform another experiment which yields another observed 
value of the statistic $\xi^{'}_{\mathrm{obs}}$; then $Pv$ is the probability 
that $\xi^{'}_{\mathrm{obs}}$ is greater than $\xi_{\mathrm{obs}}$ given that the null 
hypothesis is true. 
Namely, the $P$-value measures the strength of the evidence in support
of a null hypothesis.
  
 
\bibliographystyle{spr-mp-sola-cnd} 

\bibliography{Lognormal_Distr_Lambda}  
%
\IfFileExists{\jobname.bbl}{} {\typeout{}
\typeout{****************************************************}
\typeout{****************************************************}
\typeout{** Please run "bibtex \jobname" to obtain} \typeout{**
the bibliography and then re-run LaTeX} \typeout{** twice to fix
the references !}
\typeout{****************************************************}
\typeout{****************************************************}
\typeout{}}


\end{article} 

\end{document}